\documentclass[journal]{IEEEtran}

\usepackage{xcolor,soul,framed} 

\colorlet{shadecolor}{yellow}
\usepackage[pdftex]{graphicx}
\usepackage{microtype}
\graphicspath{{images/}}
\DeclareGraphicsExtensions{.pdf,.jpeg,.png}
\usepackage{url}

\usepackage[cmex10]{amsmath}
\usepackage{amsthm} 
\usepackage{array}
\usepackage{float}
\usepackage{mdwmath}
\usepackage{mdwtab}
\usepackage{eqparbox}
\usepackage{booktabs}
\usepackage{hyperref}
\usepackage{multirow}
\usepackage{url}
\usepackage{braket}
\usepackage{amssymb}
\usepackage[utf8x]{inputenc}
\usepackage[justification=centering]{caption}
\usepackage{quantikz} 
\usepackage{tikz}
\usepackage{adjustbox}
\usepackage[FIGTOPCAP]{subfigure}
\usepackage{subcaption}
\usepackage{mdframed}
\usepackage{bbm}
\usepackage[margin=0.7in]{geometry}

\usepackage{algorithm} 
\usepackage[noend]{algpseudocode}
\usepackage{algorithmicx}

\usepackage{mathtools}

\newtheorem{thm}{Theorem}[section]
\newtheorem{prop}[thm]{Proposition}

\newtheorem{lemma}[thm]{Lemma}

\newtheorem{rk}[thm]{Remark}

\usepackage{pythonhighlight}
\usepackage{listings}
\lstdefinestyle{lststyle}{
  commentstyle=\color{green},
  keywordstyle=\color{magenta},
  numberstyle=\tiny\color{gray},
  stringstyle=\color{purple},
  basicstyle=\ttfamily\footnotesize,
  breakatwhitespace=false,
  breaklines=true,
  captionpos=b,
  frame=lines,
  keepspaces=true,
  numbers=right,
  numbersep=5pt,
  showspaces=false,
  showstringspaces=false,
  showtabs=false,
  tabsize=2
}
\usepackage{listings, xcolor}

\definecolor{verylightgray}{rgb}{.97,.97,.97}

\lstdefinelanguage{Solidity}{
    keywords=[1]{anonymous, assembly, assert, balance, break, call, callcode, case, catch, class, constant, continue, constructor, contract, debugger, default, delegatecall, delete, do, else, emit, event, experimental, export, external, false, finally, for, function, gas, if, implements, import, in, indexed, instanceof, interface, internal, is, length, library, log0, log1, log2, log3, log4, memory, modifier, new, payable, pragma, private, protected, public, pure, push, require, return, returns, revert, selfdestruct, send, solidity, storage, struct, suicide, super, switch, then, this, throw, transfer, true, try, typeof, using, value, view, while, with, addmod, ecrecover, keccak256, mulmod, ripemd160, sha256, sha3}, 
    keywordstyle=[1]\color{blue}\bfseries,
    keywords=[2]{address, bool, byte, bytes, bytes1, bytes2, bytes3, bytes4, bytes5, bytes6, bytes7, bytes8, bytes9, bytes10, bytes11, bytes12, bytes13, bytes14, bytes15, bytes16, bytes17, bytes18, bytes19, bytes20, bytes21, bytes22, bytes23, bytes24, bytes25, bytes26, bytes27, bytes28, bytes29, bytes30, bytes31, bytes32, enum, int, int8, int16, int24, int32, int40, int48, int56, int64, int72, int80, int88, int96, int104, int112, int120, int128, int136, int144, int152, int160, int168, int176, int184, int192, int200, int208, int216, int224, int232, int240, int248, int256, mapping, string, uint, uint8, uint16, uint24, uint32, uint40, uint48, uint56, uint64, uint72, uint80, uint88, uint96, uint104, uint112, uint120, uint128, uint136, uint144, uint152, uint160, uint168, uint176, uint184, uint192, uint200, uint208, uint216, uint224, uint232, uint240, uint248, uint256, var, void, ether, finney, szabo, wei, days, hours, minutes, seconds, weeks, years}, 
    keywordstyle=[2]\color{teal}\bfseries,
    keywords=[3]{block, blockhash, coinbase, difficulty, gaslimit, number, timestamp, msg, data, gas, sender, sig, value, now, tx, gasprice, origin},   
    keywordstyle=[3]\color{violet}\bfseries,
    identifierstyle=\color{black},
    sensitive=false,
    comment=[l]{//},
    morecomment=[s]{/*}{*/},
    commentstyle=\color{gray}\ttfamily,
    stringstyle=\color{red}\ttfamily,
    morestring=[b]',
    morestring=[b]"
}

\newcommand{\C}{\mathbb C}

\newcommand{\tr}{\mathrm{tr}}

\renewcommand{\i}{\mathbf i}
\renewcommand{\j}{\mathbf j}
\renewcommand{\H}{\mathcal H} 
\renewcommand{\S}{\mathfrak S} 
\renewcommand{\SS}{\mathcal S} 
\newcommand{\F}{\mathcal F} 

\newcommand{\I}{\mathbb I}
\newcommand{\x}{X}

\newcommand{\z}{Z}

\newcommand{\cz}{\mathrm{CZ}}
\newcommand{\Cl}{\mathrm{Cl}} 

\begin{document}
\bstctlcite{IEEEexample:BSTcontrol}
    \title{Benchmarking Multipartite Entanglement Generation with Graph States}
    \author{René Zander, Colin Kai-Uwe Becker\\
    Fraunhofer Institute for Open Communication Systems (FOKUS)\\ rene.zander@fokus.fraunhofer.de, colin.kai-uwe.becker@fokus.fraunhofer.de}

\pagestyle{plain}
\pagenumbering{arabic}

\maketitle

\let\oldref\ref
\renewcommand{\ref}[1]{(\oldref{#1})}

\begin{abstract}
As quantum computing technology slowly matures and the number of available qubits on a QPU gradually increases, interest in assessing the capabilities of quantum computing hardware in a scalable manner is growing. One of the key properties for quantum computing 
is the ability to generate multipartite entangled states. In this paper, aspects of benchmarking entanglement generation capabilities of noisy intermediate-scale quantum (NISQ) devices are discussed based on the preparation of graph states and the verification of entanglement in the prepared states. Thereby, we use entanglement witnesses that are specifically suited for a scalable experiment design. This choice of entanglement witnesses can detect A) bipartite entanglement and B) genuine multipartite entanglement for graph states with constant two measurement settings if the prepared graph state is based on a 2-colorable graph, e.g., a square grid graph or one of its subgraphs. With this, we experimentally verify that a fully bipartite entangled state can be prepared on a $127$-qubit IBM Quantum superconducting QPU, and genuine multipartite entanglement can be detected for states of up to $23$ qubits with quantum readout error mitigation.
\end{abstract}
\begin{IEEEkeywords}
Quantum computing, Benchmarking, Entanglement, Entanglement Witnesses, Graph States
\end{IEEEkeywords}

\IEEEpeerreviewmaketitle

\section{Introduction}

Experiments for verifying entanglement generation capabilities of gate-based quantum computers gained traction in the recent years in line with the availability of QPUs with an increasing number of qubits, which is evident from various published results. These include showing genuine multipartite entanglement for a $27$-qubit GHZ state \cite{MWHH21ghz}, bipartite entanglement for a $65$-qubit graph state \cite{MWHH21}, genuine multipartite entanglement on $51$ qubits \cite{CWC23} and most recently analyzing bipartite and multipartite entanglement for up to $433$ qubits  \cite{KKHMH23}. Furthermore, defining a benchmarking protocol for assessing entanglement generation capabilities \cite{Hamilton22} using the volumetric benchmarking framework \cite{VolumetricFramework} was explored.

Graph states are well-known and researched due to their relevance as a universal resource state for measurement-based quantum computing \cite{OneWayQC}, and as graph codes in quantum cryptography applications and quantum error correction \cite{Hein06}. Recently, graph states were used as encoding scheme for an equivalence checking algorithm for comparing bit-strings efficiently on gate-based quantum computers \cite{SupercheQ}. 
The main focus of this paper lies in establishing graph states as a means for efficiently benchmarking and comparing aspects of entanglement generation capabilities for gate-based quantum computing architectures. For this, bipartite and genuine multipartite entanglement is detected for graph states based on a $2$-colorable graph by using different entanglement witnesses in a scalable experiment design that requires only two measurement settings. The corresponding measurement results are already sufficient to verify bipartite and multipartite entanglement not only for the entire graph state but also for states that correspond to connected subgroups of qubits. This is done by evaluating expectation values for different entanglement witnesses that are solely dependent on the obtained measurement results. This particular choice of entanglement witnesses thereby provides a new approach adding to the methods from previous publications. We demonstrate our approach through experiments on three $127$-qubit IBM Quantum QPUs. In particular, we experimentally verify that a fully bipartite entangled state can be prepared for $127$ qubits, and genuine multipartite entanglement can be detected for states of up to $23$ qubits with quantum readout error mitigation. The proposed experiments are aimed to be used as a benchmark for gate-based QPUs. The results provide fair performance comparisons for hardware architectures if the chosen graph states can be natively prepared on each QPU without limitations imposed by the respective qubit topologies.

\smallskip

The paper is structured as follows. An introduction to the structure of entanglement for multipartite systems and entanglement witnesses is provided in Section \ref{sec:entanglement}. In Section \ref{sec:graph_states}, we start with a brief overview about graph states and the stabilizer formalism, and discuss the belonging entanglement witnesses. Here, we also provide novel insights into the analysis of the structure of entanglement for subgroups of connected qubits. The experiment design, 
the obtained results, as well as aspects of benchmarking are discussed in Section \ref{sec:experiments}. Finally, we conclude with an outlook based on the experiment results in Section \ref{sec:Outlook}. The Appendix \ref{sec:Appendix} contains detailed information about the algorithmic implementation used for conducting the experiments as well as additional mathematical proofs.

\section{Entanglement}
\label{sec:entanglement}

In the following, we briefly discuss the structure of entanglement of multipartite systems as well as entanglement witnesses.
Let $M$ be a set of qubits and $m=|M|$. An $m$-qubit mixed state $\rho$ is \textit{separable} if it can be written as probabilistic mixture of separable pure states with respect to a fixed bipartition $A,B$ of the set $M$ of qubits. That is, 
\begin{equation}
\rho=\sum\limits_kp_k\rho_k^A\otimes\rho_k^B
\end{equation}
where $\rho_k^A$ and $\rho_k^B$ are pure states of the subsystems $A$ and $B$, respectively. The coefficients $p_k$ define a probability distribution, that is, they are positive and sum up to one. Denote the set of separable states by $\S$. A mixed state is \textit{bipartite entangled} if it is not separable with respect to all bipartitions of qubits of the system.

An $m$-qubit mixed state $\rho$ is \textit{fully-separable} if it can be written as probabilistic mixture of fully-separable pure states,
that is, 
\begin{equation}
\rho=\sum\limits_kp_k\rho_k^{(1)}\otimes\dotsb\otimes\rho_k^{(m)}
\end{equation}
where $\rho_k^{(i)}$ are pure states of qubit $i$.
Denote the set of fully-separable states by $\S_f$.
A mixed state is \textit{entangled} if it is not fully separable.

An $m$-qubit mixed state $\rho$ is \textit{biseparable} if it can be written as convex mixture of separable states, that is, 
\begin{equation}
\rho=\sum\limits_kq_k\rho_k
\end{equation}
where $\rho_k$ are separable states and may have different bipartitions. The coefficients $q_k$ define a probability distribution. Denote the set of biseparable states by $\S_b$. A mixed state is \textit{genuinely multipartite entangled} (GME) if it is not biseparable.
Clearly, $\S_f\subset\S\subset\S_b$.

\subsection{Entanglement witnesses}

Entanglement witnesses can serve as a helpful tool for experimentally demonstrating the presence of entanglement in a quantum state. 
An entanglement witness for genuine multipartite entanglement is an operator $W$ that has non-negative expectation on all biseparable states, i.e.,
\begin{equation}
\tr(W\rho)\geq 0,\quad\text{for all }\rho\in\S_b,
\end{equation}
and a negative expectation on at least one entangled state $\tilde{\rho}\notin\S_b$. 
That is, measuring a negative expectation for $W$ verifies that a state is genuinely multipartite entangled.
Similarly, one defines entanglement witnesses for detecting, e.g., non-full-separability (entanglement) or non-separability with respect to certain bipartitions of the set of qubits.

In an experimental setting, the prepared state deviates from the target state due to the presence of noise in state-of-the-art NISQ devices.
For a given pure state $\ket{\psi}$, the \textit{projector-based witness} \cite{TG05}
\begin{equation}
\label{eq:witness}
W=c\I-\ket{\psi}\bra{\psi}
\end{equation}
can detect genuine multipartite entanglement.
Here, $c$ is the smallest constant such that $\tr(W\rho)\geq0$ for all biseparable states $\rho\in\S_b$. 
It can be computed via Schmidt decomposition \cite{B04}.
The witness \ref{eq:witness} detects a state $\rho$ as genuinely multipartite entangled if the fidelity between $\ket{\psi}$ and $\rho$ is greater than $c$, i.e., $\F(\ket{\psi},\rho)>c$. 

In general, the number of measurement settings required to evaluate the operator \ref{eq:witness}
grows exponentially with the number of qubits.
For a scalable experiment design it is desirable to construct entanglement witnesses that require only a few 
measurement settings. For example, for $m$-qubit graph states there are witnesses of the form \cite{TG05}:
\begin{equation}
W=c_0\I-\sum\limits_{k=1}^mc_kS_k
\end{equation}
where $c_k$ are constants and the operators $S_k$ are tensor products of Pauli matrices.
Then each $S_k$ requires only one measurement setting, so that $W$ can be evaluated with at most $m$ measurement settings.

\section{Graph states}
\label{sec:graph_states}

\newcommand{\stab}{\mathrm{s}}
\newcommand{\V}{\mathcal{V}}
\newcommand{\tol}{\mathrm{tol}}


We provide a brief introduction to graph states \cite{Hein06,JMG11}.

Let $G=(V,E)$ be a graph where $V$ is the set of vertices and $E$ is the set of edges. Denote the number of vertices $|V|$ by $n$.
A graph state $\ket{G}\in(\C^2)^{\otimes n}$ can be associated as follows: vertices represent qubits initialized in the $\ket{+}=(\ket{0}+\ket{1})/\sqrt{2}$ state, 
and edges $e=(i,j)$ represent the controlled $\z$-operation $\cz^{(i,j)}$ acting on qubits $i$ and $j$. 
Recall that 
\begin{equation}
\label{eq:CZ}
\cz^{(i,j)}=\pi_{+}^i\otimes\I^j+\pi_{-}^i\otimes\sigma_z^j
\end{equation}
where $\pi_{\pm}^i=(\I\pm\sigma_z^i)/2$ 
are the projectors onto the eigenspaces of the operator $\sigma_z^i$ for eigenvalues $+1$ and $-1$, respectively.

That is, the \textit{graph state} $\ket{G}$ is defined as 
\begin{equation}
\label{eq:graph_state}
\ket{G}=\prod_{(i,j)\in E}\cz^{(i,j)}\ket{+}^{\otimes n}.
\end{equation}
Define the operators
\begin{equation}
\label{eq:stabilizer}
S_i=\sigma_x^i\prod\limits_{j\in N_i}\sigma_z^j,\quad i=1,\dotsc,n,
\end{equation}
where $N_i=\{j\in V\mid (i,j)\in E\}$ is the set of neighbors of the vertex $i$. 
The operators $S_i$ commute and generate a set of so-called \textit{stabilizer operators} that consists of $2^n$ elements, 
\begin{equation}
\SS=\left\{\prod\limits_{i=1}^nS_i^{x_i}\mid x\in\{0,1\}^n\right\}.
\end{equation} 
The graph state $\ket{G}$ is the unique state that is an eigenstate to eigenvalue $+1$ for all $S_i$, that is,
\begin{equation}
S_i\ket{G}=\ket{G},\quad\text{for all } i=1,\dotsc,n.
\end{equation}
The projector on $\ket{G}$ can be written as product of so-called \textit{stabilizer projectors} 
$(\I+S_i)/2$ onto the eigenspace of $S_i$ with eigenvalue $+1$, that is, 
\begin{equation}
\ket{G}\bra{G}=\prod\limits_{i\in V}\dfrac{\I+S_i}{2}.
\end{equation}

For a graph $G=(V,E)$ and a subset $U\subset V$, we define the 
stabilizer projector of the subset $U$ as
\begin{equation}
P(U,G)=\prod\limits_{i\in U}\frac{\I+S_i}{2}.
\end{equation}
In particular, $\ket{G}\bra{G}=P(V,G)$.

\subsection{Entanglement witnesses for bipartite entanglement}

Here, we discuss how to detect bipartite entanglement in a prepared quantum state $\rho$ with target quantum state $\ket{G}$.

The following entanglement witness can be used to detect non-separability. It is known that the same witness can be used to rule out full separability \cite{TG05,Hein06}. Here, we generalize this result to the weaker assumption of separability.

\begin{prop}
\label{prop:bipartite_entanglement}
Let $G=(V,E)$ be a graph and $(i,j)\in E$.
The operator $W_{ij}$ can witness non-separability (entanglement),
\begin{equation}
W_{ij}=\I-S_i-S_j
\end{equation}
with $\langle W_{ij}\rangle\geq 0$ for all states $\rho\in (\C^2)^{\otimes n}$ that are separable with respect to any bipartition $A, B$ with $i\in A$ and $j\in B$.

\begin{proof}
This is a reformulation of the necessary condition for separability given in Proposition \ref{prop:separability_graph}.
\end{proof}
\end{prop}

Let $G=(V,E)$ and $\rho\in (\C^2)^{\otimes n}$ be a density operator with qubits $V$. For an edge $e=(i,j)$ we define its weight $w(e)=\langle W_{ij}\rangle$.
Let $G'=(V,E')$ with $E'=\{e\in E\mid w(e)<0\}$ be the subgraph of $G$ with all edges deleted that have non-negative weight. Then the connected components of $G'$ correspond to bipartite entangled subsets of qubits.

\subsection{Entanglement witnesses for multipartite entanglement}

In the following, we discuss how to detect multipartite entanglement in a prepared quantum state $\rho$ with target quantum state $\ket{G}$.

For any graph state $\ket{G}$ the projector-based witness
\begin{equation}
\label{eq:witness_gs}
W(G)=\frac12\I-\ket{G}\bra{G}
\end{equation}
can detect genuine multipartite entanglement, 
with $\langle W(G)\rangle\geq0$ for all biseparable states.
This follows from the fact that for any graph state the fidelity between $\ket{G}$ and any biseparable state $\rho$ is
upper bounded by $1/2$ \cite{ZZYM19}. It was also shown that this bound is tight in the sense that there is a
biseparable state $\rho$ such that $\tr(\ket{G}\bra{G}\rho)=1/2$.

To measure the projector $\ket{G}\bra{G}$, one considers the expansion 
\begin{equation}
\ket{G}\bra{G}=\frac{1}{2^n}\sum_{S\in\SS}S
\end{equation}
which is a weighted sum of all $2^n$ stabilizer operators.
Therefore, the number of stabilizer measurements grows exponentially in the number of qubits and is practically feasible only for small systems. 

With Lemma \ref{lem:operators} we obtain entanglement witnesses that may require fewer measurements \cite{ZZYM19}.
Let $\V=\{V_1,\dotsc,V_k\}$ be a partition of the vertex set $V$ into disjoint subsets. Then the operator	
\begin{equation}
\label{eq:witness_gs_gen}
W(\V,G)=\left(k-\frac12\right)\I-\sum\limits_{l=1}^kP(V_l,G)
\end{equation}
can detect genuine multipartite entanglement,  
with $\langle W(\V,G)\rangle\geq0$ for all biseparable states.
Typical choices of the partition $\V$ are the following:
if $\V=\{V\}$, we obtain the projector-based witness $W(\{V\},G)=W(G)$. 
If $\V=\{\{i\}\}_{i\in V}$, we obtain (up to a factor of $1/2$) the \textit{stabilizer sum witness} \cite{TG05,Hamilton22}
\begin{equation}
W^{\stab}(G)=(n-1)\I-\sum\limits_{l=1}^nS_l.
\end{equation}
Each stabilizer $S_l$ is a tensor product of Pauli matrices 
and hence requires only one measurement setting. Then $W^{\stab}(G)$ can be evaluated with at most $n$ measurement settings.

Lastly, a map $c\colon V\rightarrow C$, where $C$ is the set of colors, is a \textit{proper vertex coloring} if any two vertices that have the same color are not connected by an edge. A graph is called \textit{$k$-colorable} if it has a coloring with $|C|=k$ colors.
The minimal number $k$ of colors is called the \textit{chromatic number} $\chi$ of the graph.
A coloring induces a partition $\V^{c}=\{V_c\}_{c\in C}$ of the vertex set $V$ into disjoint subsets such that any two vertices in the same subset are not connected by an edge. The case where the partition $\V$ corresponds to such a proper vertex coloring was investigated for GHZ and $1$-D cluster states \cite{TG05} and subsequently proposed as a systematic method for construction 
of entanglement witnesses for graph states \cite{ZZYM19}. We denote the \textit{coloring-based witness} by $W^c(G)=W(\{V_c\}_{c\in C},G)$.
In this case, the expectation of each projector $P(V_c,G)$ can be computed with one measurement setting $\otimes_{i\in V_c}\x_i\otimes_{j\in V\setminus V_c}\z_j$. Then the computation of the expectation of $W^c(G)$ requires only $|C|$ measurement settings.
In particular, for graph states corresponding to $2$-colorable graphs, e.g., $1$-D and $2$-D cluster states, entanglement witnesses that require only two measurement settings can be found.

\smallskip

In experiments, the prepared state $\rho$ differs from the target graph state $\ket{G}\in (\C^2)^{\otimes n}$
due to the presence of noise. 
The white noise tolerance is commonly used as indicator of the robustness of a witness \cite{JMG11}.
It is defined as follows. 
For a state $\tilde{\rho}$ and a witness $W$, consider the state 
\begin{equation}
\label{eq:white_noise}
\rho(p)=(1-p)\tilde{\rho}+p\I/2^n,
\end{equation}
for $p\in [0,1]$, that is, $\rho(p)$ is a stochastic mixture of the state $\tilde{\rho}$ and the maximally mixed state.
Then the \textit{white noise tolerance} is the maximal $p_{\tol}$ such that $\rho(p)$ is detected by the witness $W$, i.e., $\tr(W\rho(p))<0$ for all $p\in [0,p_{\tol})$.
For the witnesses $W(\V,G)$, for some partition of the vertex set $\V=\{V_1,\dotsc,V_k\}$, 
we have $1/k\geq p_{\tol}>1/(2k)$ \cite{ZZYM19}. 
In fact, for certain graph states, e.g., $1$-D and $2$-D cluster states, one can construct witnesses 
such that their white noise tolerance approaches one as the number of qubits increases.
That is, under the presence of white noise the fidelity between the prepared state $\rho$ and the target graph state $\ket{G}$ can
decrease exponentially with the number of qubits, but the state $\rho$ is still genuinely multipartite entangled and can be detected by a witness \cite{JMG11}. Yet, as these witnesses are an augmentation of the projector-based witness $W(G)$, the number of local measurement settings grows exponentially with the number of qubits.

Finally, a partition $\widetilde{\V}=\{\widetilde{V}_1,\dotsc,\widetilde{V}_l\}$ is a refinement of the partition $\V=\{V_1,\dotsc,V_k\}$ if for all $i\in\{1,\dotsc,l\}$ there is a $j\in\{1,\dotsc,k\}$ such that $\widetilde{V}_i\subset V_j$. Then with Lemma \ref{lem:operators}, we see that 
\begin{equation}
\label{eq:witness_refinement}
W(\widetilde{\V},G)\geq W(\V,G).
\end{equation}
Here, for Hermitian operators $A,B$, we write $A\geq B$ indicating that $(A-B)$ is positive semidefinite.
In particular, the witness $W(\widetilde{\V},G)$ has a lower white noise tolerance. 
In Appendix \ref{sec:implementation}, it is shown that considering witnesses corresponding to refinements of a partition $\V$ can still be useful, specifically in the context of quantum readout error mitigation.

\subsection{Entanglement witnesses for subgraphs}

\newcommand{\EE}{\widetilde{E}}


Given sampled measurement results corresponding to a prepared state $\rho$ with target graph state $\ket{G}$,
we discuss how to obtain information on the ability of the QPU to generate multipartite entangled 
states that correspond to subgraphs $G'\subset G$.

\begin{figure}[h]
\centering
\begin{tikzpicture}[scale=0.6]
	\tikzstyle{vertex}=[shape=circle,draw=black]
	\node[vertex] (v1) at (0,0) {1};
	\node[vertex] (v2) at (2,0) {2};
	\node[vertex] (v3) at (4,0) {3};
	\node[vertex] (v4) at (0,-2) {4};
	\node[vertex] (v5) at (2,-2) {5};
	\node[vertex] (v6) at (4,-2) {6};
	\node[vertex] (v7) at (0,-4) {7};
	\node[vertex] (v8) at (2,-4) {8};
	\node[vertex] (v9) at (4,-4) {9};
	\node[vertex] (v10) at (6,0) {\textcolor{white}{1}};
	\node[vertex] (v11) at (6,-2) {\textcolor{white}{1}};
	\node[vertex] (v12) at (6,-4) {\textcolor{white}{1}};
	\tikzstyle{auxiliary}=[shape=circle,draw=white,scale=0.1]
	\node[auxiliary] (a1) at (-1,-3) {};
	\node[auxiliary] (a2) at (3,-3) {};
	\node[auxiliary] (a3) at (3,1) {};
		
	\path[-] (v1) edge[thick] node {} (v2);
	\path[-] (v2) edge[thick,color=red] node {} (v3);
	\path[-] (v4) edge[thick] node {} (v5);
	\path[-] (v5) edge[thick,color=red] node {} (v6);
	\path[-] (v7) edge[thick] node {} (v8);
	\path[-] (v8) edge[thick] node {} (v9);
		
	\path[-] (v1) edge[thick] node {} (v4);
	\path[-] (v2) edge[thick] node {} (v5);
	\path[-] (v3) edge[thick] node {} (v6);
	\path[-] (v4) edge[thick,color=red] node {} (v7);
	\path[-] (v5) edge[thick,color=red] node {} (v8);
	\path[-] (v6) edge[thick] node {} (v9);
	
	\path[-] (a1) edge[color=green,style=dashed] node {} (a2);
	\path[-] (a2) edge[color=green,style=dashed] node {} (a3);
	
	\path[-] (v3) edge[thick] node {} (v10);
	\path[-] (v6) edge[thick] node {} (v11);
	\path[-] (v9) edge[thick] node {} (v12);
	\path[-] (v10) edge[thick] node {} (v11);
	\path[-] (v11) edge[thick] node {} (v12);
\end{tikzpicture}
\caption{The $2$-D cluster graph $\Cl_{3\times 4}$. Consider the subgraph $G'=(V',E')$ with $V'=\{1,2,4,5\}$ and $E'=\{(1,2),(1,4),(2,5),(4,5)\}$. Then $N(V')=\{3,6,7,8\}$ and, as indicated by the red lines, $\EE=\{(2,3),(4,7),(5,6),(5,8)\}$.}
\label{fig:subgraph}
\end{figure}
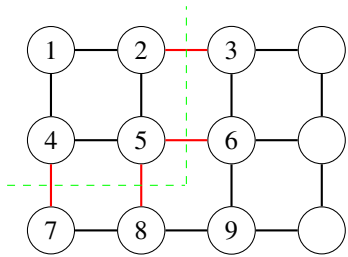

Let $G=(V,E)$ be a graph and $G'=(V',E')\subset G$ be a subgraph with $E'=\{(i,j)\in E\mid i,j\in V'\}$. That is, $G'$ is the subgraph induced by the subset of vertices $V'\subset V$. Let the neighborhood $N(V')\subset V$ be the subset of all vertices in $G\setminus G'$ that are adjacent to at least one vertex in $G'$.
Let $\EE\subset E$ be the subset of edges that connect $G'$ with $G\setminus G'$. This is illustrated in Figure \ref{fig:subgraph}.

Consider the stabilizer projectors
\begin{eqnarray}
P(V',G)&=&\prod\limits_{v\in V'}\frac12(\I+S_v),\\
P(V',G')&=&\prod\limits_{v\in V'}\frac12(\I+S_v')
\end{eqnarray}
where $S_v$ and $S_v'$ are the stabilizers of $\ket{G}$ and $\ket{G'}$, respectively.
Clearly, $P(V',G')=\ket{G'}\bra{G'}$. The operator $P(V',G')$ is obtained from $P(V',G)$ by replacing all Pauli operators 
acting on the qubits in $G\setminus G'$ with identities.

One could aim to compute the expectation of the projector-based witness $W(\{V'\},G')$ with respect to the graph state $\ket{G}$.
For example, consider the $1$-D cluster state $\rho=\ket{G}\bra{G}$ on four qubits defined by the graph $G=(V,E)$ with 
$V=\{0,1,2,3\}$, $E=\{(0,1),(1,2),(2,3)\}$, and let $G'=(V',E')$ with $V'=\{1,2\}$, $E'=\{(1,2)\}$ be a subgraph. In this case, a straightforward computation shows that 
the reduced density operator $\rho^{\{1,2\}}$ of the subsystem of qubits $\{1,2\}$ is given by 
$\rho^{\{1,2\}}=\tr_{\{0,3\}}\rho=(1/4)\I_2\otimes\I_2$. That is, the reduced state $\rho^{\{1,2\}}$ is separable!
Accordingly, we have $\langle W(V',G')\rangle\geq 0$.

More generally, given an entangled state $\rho$ on qubits $V$, the reduced state $\rho^{V'}$ with respect to the subset of qubits $V'\subset V$ might not be entangled.

\smallskip

Instead, we propose measuring the projector $P(V',G)$ to obtain information 
on the ability of the QPU to generate a multipartite entangled 
state that corresponds to the subgraph $G'\subset G$.
An interpretation of $P(V',G)$ is given by the following result.

\begin{lemma}
\label{lem:projector_subsystem}
The following identity holds:
\begin{equation}
\label{eq:projectors_cz}
P(V',G)=\prod\limits_{(i,j)\in \EE}CZ^{(i,j)}\ket{G'}\bra{G'}\otimes\I\prod\limits_{(i,j)\in\EE}CZ^{(i,j)}.
\end{equation}

\begin{proof}
Recall that $\ket{G'}\bra{G'}=\prod_{v\in V'}(\I+S_v')/2$.
Let $(i,j)\in\EE$. The unitary $\cz^{(i,j)}$ commutes with all stabilizer operators $S_v'$ for $v\notin\{i,j\}$.
There is exactly one vertex $v\in V'$ such that $v\in\{i,j\}$.
Without loss of generality, let $v=i$. Then using \ref{eq:CZ} and the relation $\sigma_x\pi_{\pm}=\pi_{\mp}\sigma_x$, we
find 
\begin{equation}
\cz^{(i,j)}S_i'\cz^{(i,j)}=\cz^{(i,j)}(\pi_-^i\otimes\I^j+\pi_+^i\otimes\sigma_z^j)S_i'=\sigma_z^jS_i'.
\end{equation}
The last equation follows from a straightforward calculation using the identities
$\pi_{\pm}^2=\pi_{\pm}$, $\pi_{+}\pi_{-}=\pi_{-}\pi_{+}=0$ and $\pi_{+}+\pi_{-}=\I$ for the projectors $\pi_{\pm}$.
Then the claim follows by induction on the set of edges $\EE$.
\end{proof}
\end{lemma}

\begin{prop}
\label{prop:projector_subsystem}
The following identity holds:
\begin{equation}
\tr(P(V',G)\rho)=\tr(\ket{G'}\bra{G'}\hat{\rho}^{V'})
\end{equation}
where 
\begin{equation*}
\hat{\rho}=\prod\limits_{(i,j)\in \EE}CZ^{(i,j)}\rho\prod\limits_{(i,j)\in \EE}CZ^{(i,j)}.
\end{equation*}

\begin{proof}
With equation \ref{eq:projectors_cz} and the cyclic property of the trace we find 
$\tr(P(V',G)\rho)=\tr(\ket{G'}\bra{G'}\otimes\I\hat{\rho})$. Then the claim follows from
the properties of the partial trace.
\end{proof}
\end{prop}

That is, the state $\hat{\rho}$ is obtained from the state $\rho$ by applying controlled-$\z$ operations to all
pairs of qubits $(i,j)\in\EE$. If $\rho=\ket{G}\bra{G}$, this amounts to removing all edges connecting $G'$ with $G\setminus G'$.
Then the resulting graph state $\hat{\rho}=\ket{G'}\bra{G'}\otimes\ket{G\setminus G'}\bra{G\setminus G'}$
is a product of the graph states for the two subgraphs. In this case, we have $\tr(\ket{G'}\bra{G'}\hat{\rho})=1$.

In the language of entanglement witnesses this can be stated as follows:

\begin{prop}
\label{prop:witness_subsystem}
Let $\V'$ be a partition of the vertex set $V'$.
The operator $W(\V',G)$ as defined in \ref{eq:witness_gs_gen} can detect genuine multipartite entanglement, with $\langle W(\V',G)\rangle\geq0$ 
for all states $\rho$ on the system $V$ such that $\hat{\rho}^{V'}$ is biseparable.
\begin{proof}
For $\V'=\{V'\}$ this is a consequence of Proposition \ref{prop:projector_subsystem} and the fact that $W=\frac12\I-\ket{G'}\bra{G'}$ is an entanglement witness for $\ket{G'}$. Then the claim for arbitrary partitions follows from Lemma \ref{lem:operators}.
\end{proof}
\end{prop}

Let $G_1,G_2\subset G$ be disjoint subgraphs such that $G_1\cup G_2$ is connected, and let $\V_1,\V_2$ be partitions of the set
of vertices $V_1,V_2$ of the subgraphs $G_1,G_2$, respectively. Then we have
\begin{equation}
\label{eq:witness_partition}
W(\V_1\cup \V_2,G)=\frac12\I+W(\V_1,G)+W(\V_2,G).
\end{equation}

In experiments, the state $\hat{\rho}^{V'}$ differs from $\ket{G'}\bra{G'}$ due to the presence of noise. In particular,
it is also affected by non-local noise acting on qubits in the neighborhood $N(V')$. Therefore, we assume that
$\langle W(\V',G)\rangle_{\rho}\geq\langle W(G')\rangle_{\rho'}$ where $\rho$ and $\rho'$ are the prepared states corresponding to the graph states $\ket{G}$ and $\ket{G'}$, respectively. 
In this sense, evaluating an entanglement witness $W(\V',G)$ on the prepared state $\rho$ yields information on the ability of the QPU to prepare the graph state $\ket{G'}$.

\section{Experiments}
\label{sec:experiments}

\subsection{Experiment Design}

\subsubsection{State preparation}

\begin{figure}[h]
\centering
\includegraphics[scale=0.55]{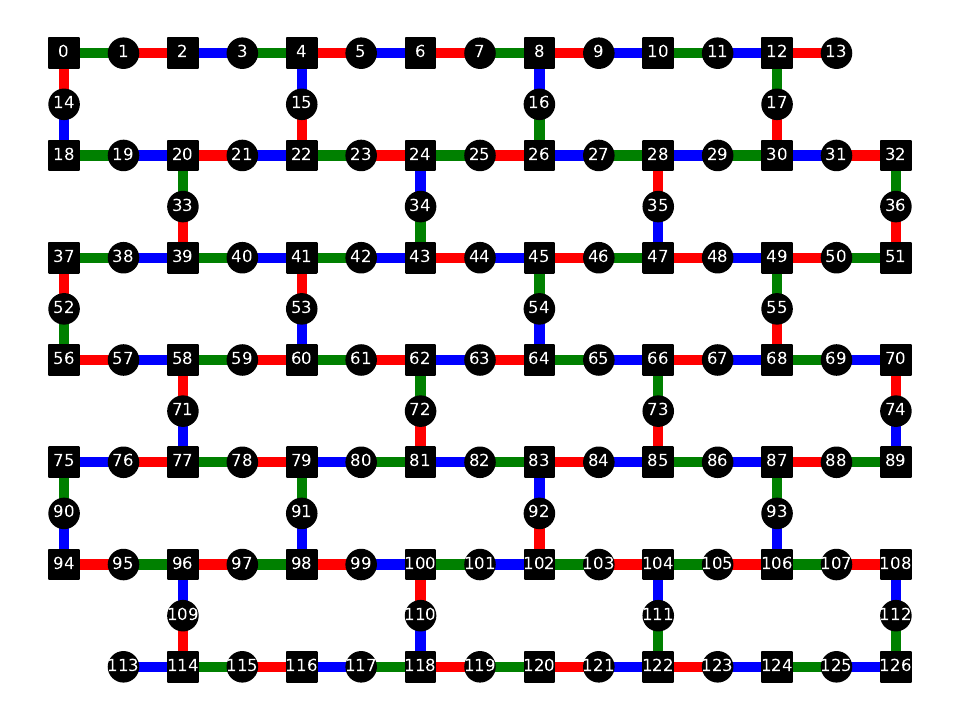}
\caption{A visualization of the heavy-hex layout of the $127$-qubit IBM devices. The edge coloring (red, green, blue) corresponds to three layers of controlled-$Z$ gates. Within each layer, the controlled-$Z$ gates can be executed in parallel. The shape of the nodes (round, rectangular) corresponds to a vertex coloring of the graph.}
\label{fig:design}
\end{figure}

We prepare the native graph state $G=(V,E)$, i.e., the graph state corresponding to the graph defined by the coupling map of the device, on the $127$-qubit IBM Quantum superconducting devices \texttt{ibm\_brisbane}, \texttt{ibm\_sherbrooke} and \texttt{ibm\_cusco}. All three devices have the same so-called heavy-hex layout, as shown in Figure \ref{fig:design}. 
All qubits are prepared in the $\ket{+}=(\ket{0}+\ket{1})/\sqrt{2}$ state by applying a Hadamard gate to their initial $\ket{0}$ state. Then the controlled-$Z$ gates corresponding to the edges are applied in three layers. Within each layer, the controlled-$Z$ gates are executed in parallel.

\subsubsection{Measurements}

Since the heavy-hex graph is $2$-colorable, we measure the prepared graph state in two measurement settings $\otimes_{i\in V_1}\x_i\otimes_{j\in V_2}\z_j$ and $\otimes_{i\in V_2}\x_i\otimes_{j\in V_1}\z_j$, where $\{V_1,V_2\}$ is the partition of the vertex set $V$ corresponding to the coloring as indicated in Figure \ref{fig:design}. For each measurement setting, $N=30000$ shots are executed. In the following, these measurement results are used to calculate expectations of stabilizers projectors and with this, entanglement witnesses for bipartite and multipartite entanglement.

\subsubsection{QREM}
\label{sec:qrem}
\newcommand{\noisy}{\mathrm{noisy}}
\newcommand{\ideal}{\mathrm{ideal}}

Quantum readout error mitigation aims to correct measurement errors by a classical post-processing of the measurement outcomes \cite{Bravyi21,M3}.
Measurement noise for a system of $M$ qubits can be characterized classically by the relation
\begin{equation}
\label{eq:qrem}
p_{\noisy}=A\cdot p_{\ideal}
\end{equation}
where $p_{\noisy}$ is the $2^M$-dimensional probability vector describing the distribution of the measurement outcomes in the presence of measurement errors, and $p_{\ideal}$ is the $2^M$-dimensional 
probability vector describing the distribution of measurement outcomes in the absence of measurement errors (but still including, e.g., gate errors), and $A$ is a $2^M\times 2^M$-dimensional stochastic matrix. The entry $A_{ij}$ is the probability of observing the outcome $i\in\{0,\dotsc,2^M-1\}$ provided that the ideal outcome is $j\in\{0,\dotsc,2^M-1\}$.
Then equation \ref{eq:qrem} can be solved for $p_{\ideal}$. Note that the result is not necessarily a probability distribution but a quasiprobability distribution: it may contain negative values but still sums up to one. This quasiprobability distribution can be used to compute an unbiased estimate for the expectation of an observable \cite{Bravyi21}.

In the tensor product noise model \cite{Bravyi21}, we assume that the noise acts independently on each qubit, i.e.,
\begin{equation}
A=\bigoplus\limits_{k=0}^{M-1}A^{(k)}.
\end{equation}
Here, $A^{(k)}$ is the calibration matrix for qubit $k$ in the computational basis, defined as  
\begin{equation}
A^{(k)}=\begin{pmatrix}
1-P_{0,1}^{(k)}&P_{1,0}^{(k)}\\
P_{0,1}^{(k)}&1-P_{1,0}^{(k)}
\end{pmatrix}
\end{equation}
where $P_{i,j}^{(k)}$ is the probability of measuring qubit $k$ in state $i\in\{0,1\}$ if the prepared state is $j\in\{0,1\}$. The error rates $p_{i,j}^{(k)}$ are obtained from $\mathcal O(M)$ calibration circuits.

In general, this error mitigation method scales only to a small number of qubits $M$. However, it can be utilized for large systems when expectations of $m$-local observables (for a small number $m$) are computed.
Recall that an observable is $m$-local if it can be decomposed as $\sum_lO_l$ where each term $O_l$ is a Hermitian operator acting on at most $m$ qubits. 
In this case, the expectation for each observable $O_l$ can be computed from the marginal distribution with respect to at most $m$ qubits. 
The $2^m\times 2^m$-dimensional calibration matrices for mitigating the marginal distributions are the tensor products of the calibration matrices for the respective qubits.
We apply this method to calculate mitigated expectations of stabilizer projectors and thereby entanglement witnesses.
In particular, this approach is suitable for evaluating the stabilizer sum witness for graph states if the belonging graph has a low maximum vertex degree, e.g., for heavy-hex graphs.
It may occur that the readout error mitigation yields non-physical values $\braket{P(U,G)}>1$. Therefore, we cap the expectations of stabilizer projectors at $1$.
Details on the implementation of evaluating entanglement witnesses with the described readout error mitigation method are given in Appendix \ref{sec:implementation}.

\subsection{Results}
\label{sec:experiments_results}

We evaluate bipartite entanglement witnesses, and multipartite entanglement witnesses for subgraphs.

\subsubsection{Bipartite entanglement}

We compute expectations of the bipartite entanglement witnesses 
\begin{equation}
\label{eq:experiments_W_bi}
W_{ij}=\I-S_i-S_j
\end{equation}
for all edges $e=(i,j)$ in the graph $G$. 
Negative expectations $\langle W_{ij}\rangle<0$ show that the system is not separable with respect to the pair of qubits $i$ and $j$. That is, there is no bipartition $A, B$ with $i\in A$, $j\in B$ of the set of qubits $V$ such that the prepared state is separable with respect to the bipartition $A, B$.
The connected subgraphs induced by the edges with negative expectations correspond to bipartite entangled regions of the device. 

The results are illustrated in Figures \ref{fig:brisbane}, \ref{fig:sherbrooke} and \ref{fig:cusco} for the devices \texttt{ibm\_brisbane}, \texttt{ibm\_sherbrooke} and \texttt{ibm\_cusco}, respectively. Notably, for \texttt{ibm\_brisbane} full $127$-qubit bipartite entanglement can be detected when QREM is applied.

Finally, note that similar results on bipartite entanglement were presented for the (now retired) devices \texttt{ibmq\_rochester} ($52$ qubits) and \texttt{ibmq\_manhattan} ($65$ qubits) \cite{MWHH21}, and most recently also for \texttt{ibm\_washington} ($127$ qubits) and \texttt{ibm\_seattle} ($433$ qubits) \cite{KKHMH23}. Information on bipartite entanglement was obtained by performing full quantum state tomography (QST) on every pair of connected qubits and their nearest neighbors, and then computing the negativity between every pair of connected qubits. In general, QST on $n$ qubits requires $3^n$ measurement settings.
If QST is performed for each pair of connected qubits, the total number of measurement settings scales linearly in the number of these pairs.
As shown recently, this scaling can be reduced to a constant factor by performing QST in parallel \cite{KKHMH23}. In contrast, in this work we show that bipartite entanglement can be characterized by measuring the prepared graph state in only two measurement settings (for $2$-colorable graphs) and calculating the bipartite entanglement witnesses \ref{eq:experiments_W_bi}.

\smallskip

\subsubsection{Multipartite entanglement}

We compute expectations of multipartite entanglement witnesses with respect to subgraphs $G'=(V',E')\subset G$.
Denote the number of vertices $|V'|$ by $n'$.
The following entanglement witnesses can be evaluated with only two measurement settings:

(i) the stabilizer sum witnesses (SSW)
\begin{equation}
\label{eq:experiments_W_stab}
W^{\stab}(G',G)=(n'-1)\I-\sum\limits_{l=1}^{n'}S_l.
\end{equation}
The main advantage in utilizing this witness is that it can be computed efficiently by summing up the previously calculated expectations of the stabilizers for the qubits in $V'$. 
However, it comes with a theoretical disadvantage of having the lowest white noise tolerance $p_{\tol}=1/n'$ among all witnesses of the form \ref{eq:witness_gs_gen}. 

Here, the SSW is utilized as follows.

\smallskip

First, we evaluate the SSW for all subgraphs $G'\subset G$ that are isomorphic to the $1$-D cluster graph $\Cl_n$, for $n=2,\dotsc,30$. If $\braket{W^{\stab}(G',G)}<0$ for a subgraph $G'$, this indicates that the graph state $\ket{G'}$ can be prepared on the device and verified as GME. In the following, we say that the state $\ket{G'}$ can be verified as GME.
For each number of qubits $n$, we identify the subgraph $G^*_n\subset G$ that minimizes the expectation $\braket{W^{\stab}(G',G)}$ over all subgraphs $G'$ that are isomorphic to $\Cl_n$. Expectations are calculated with and without QREM. 

\begin{figure*}[htbp]
\centering
\subfigure[\texttt{ibm\_brisbane}]{
\label{fig:cluster_brisbane}
\includegraphics[scale=0.55,trim={0.5cm 0cm 0cm 0cm}]{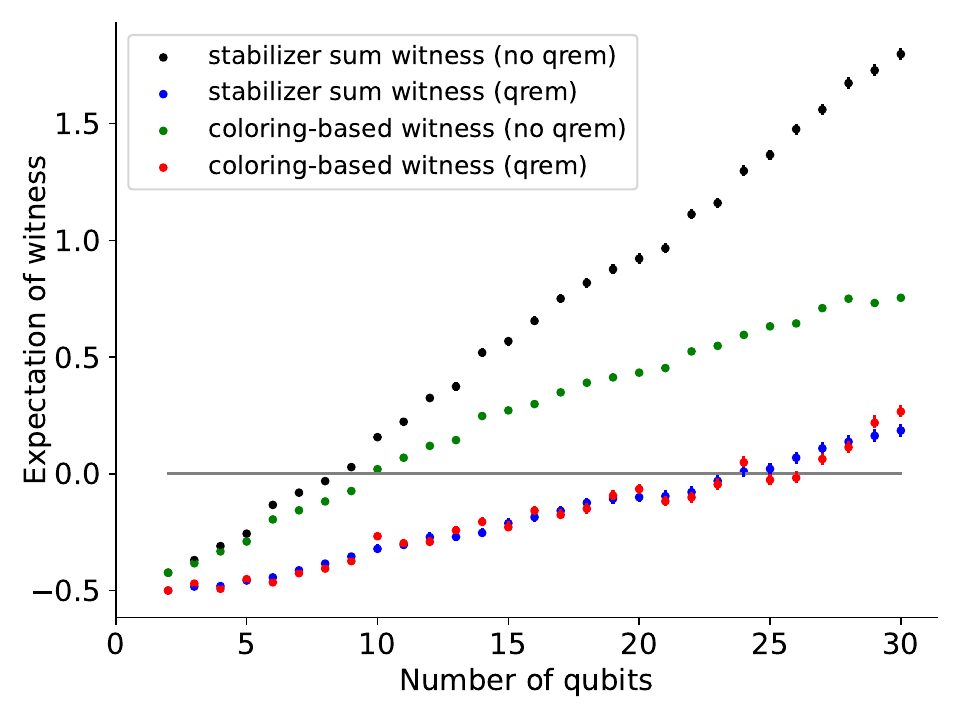}
}
\subfigure[\texttt{ibm\_sherbrooke}]{
\label{fig:cluster_sherbrooke}
\includegraphics[scale=0.55,trim={0.5cm 0cm 0cm 0cm}]{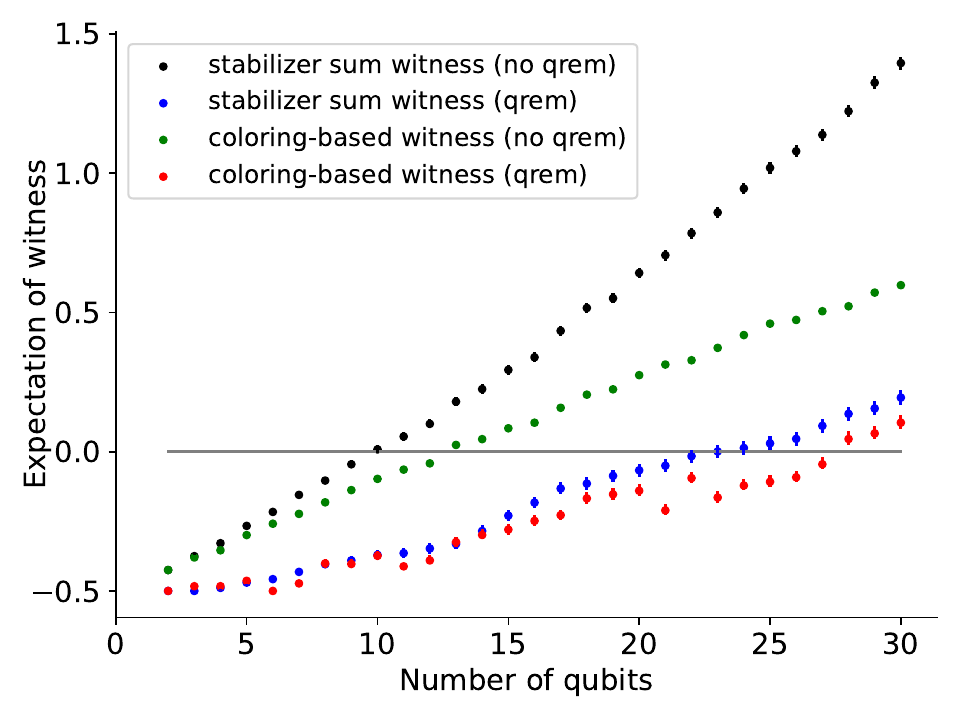}
}

\subfigure[\texttt{ibm\_cusco}]{
\label{fig:cluster_cusco}
\includegraphics[scale=0.55,trim={0.5cm 0cm 0cm 0cm}]{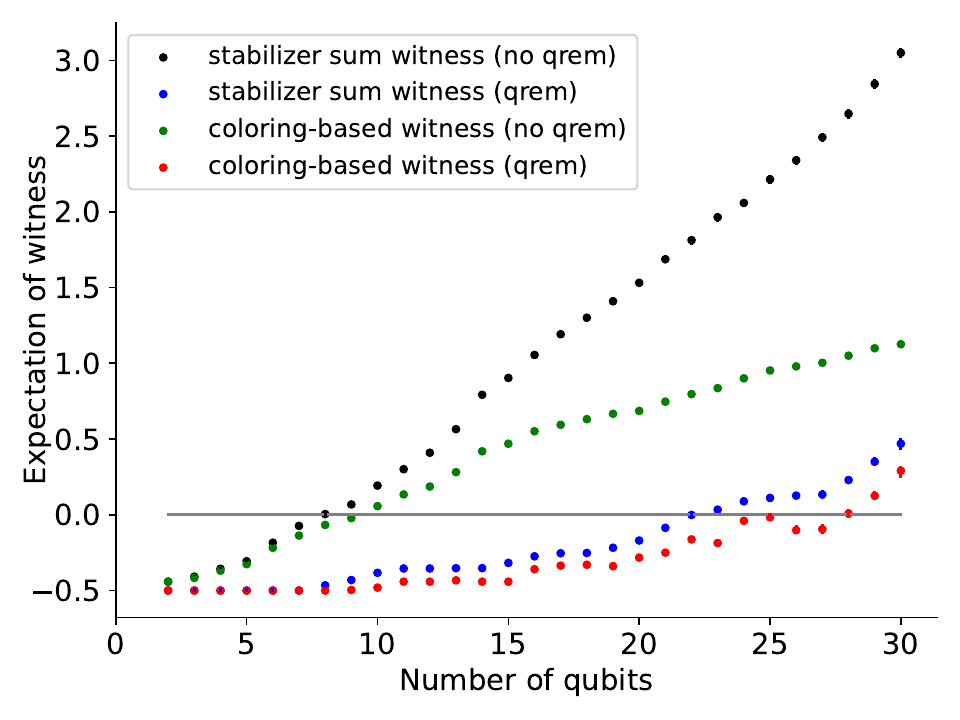}
}
\caption{Minimal expectations of the SSW (up to a factor of $1/2$) over all subgraphs that are isomorphic to the $1$-D cluster graph $\Cl_n$, for $n=2,\dotsc,30$. Expectations are computed without QREM (black) and with QREM (blue). Expectations of the CBW (green) and a refinement of the CBW (red) for the subgraphs that minimize the SSW without QREM and with QREM, respectively. GME can be verified if expectations are less than zero. The $95\%$ confidence intervals are computed with bootstrapping methods. Note that due to the large sample size $N=30000$ error bars are barely visible.}
\label{fig:cluster}
\end{figure*}

The results are illustrated in Figure \ref{fig:cluster}.
For the devices \texttt{ibm\_brisbane}, \texttt{ibm\_sherbrooke} and \texttt{ibm\_cusco}, the results indicate that a $23$-qubit, $21$-qubit and $21$-qubit $1$-D cluster state can be verified as genuinely multipartite entangled when QREM is applied, respectively.

\smallskip

Secondly, we calculate the SSW for all $12$-qubit heavy-hex unit cells in the graph. Expectations are calculated with and without QREM.

\begin{figure}[htbp]
\centering
\subfigure[\texttt{ibm\_brisbane} (no QREM)]{
\label{fig:brisbane_no_qrem}
\includegraphics[scale=0.55]{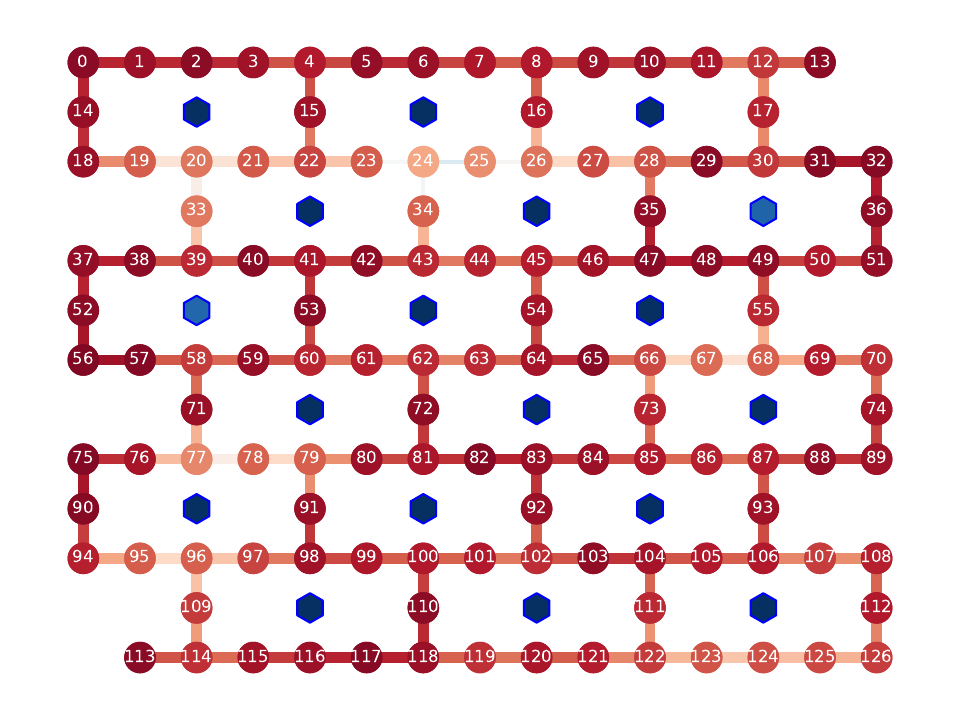}
}
\subfigure[\texttt{ibm\_brisbane} (QREM)]{
\label{fig:brisbane_qrem}
\includegraphics[scale=0.55]{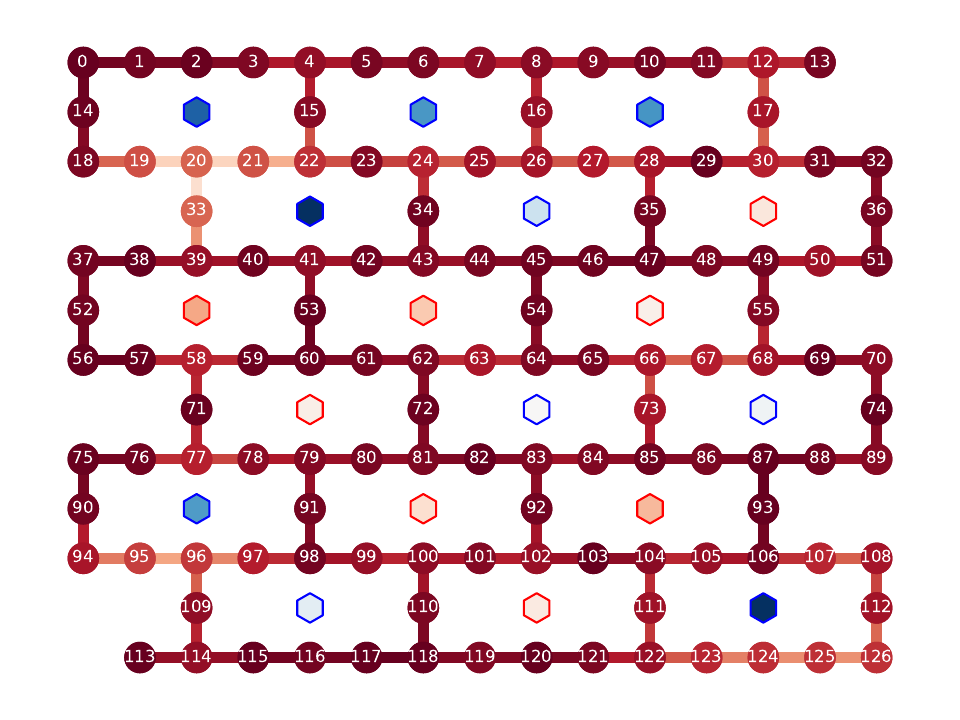}
}
\subfigure{
\includegraphics[scale=0.5]{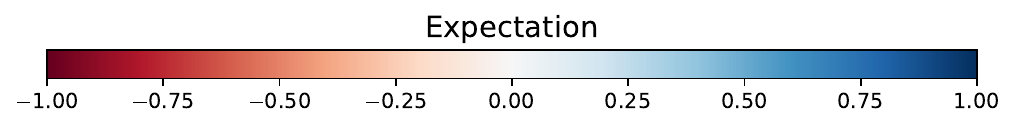}
}
\caption{A visualization of entanglement in the native graph state on the $127$-qubit \texttt{ibm\_brisbane} device. The color of each node $i$ represents the expectation of the stabilizer $-S_i$. The color of each egde $(i,j)$ represents the expectation of the entanglement witness $W_{ij}$. Thick (red) edges indicate that the system is non-separable with respect to the pair of qubits $(i,j)$ with $95\%$ confidence, and thin (blue) edges indicate that non-separability was not detected with confidence. The connected subgraphs induced by the thick (red) edges correspond to bipartite entangled regions of the device. The largest bipartite entangled regions consist of (a) $125$ qubits, and (b) $127$ qubits.
The color of the hexagons represent the expectation (capped at $1$) of the stabilizer sum witness \ref{eq:experiments_W_stab} for the corresponding heavy-hex unit-cell. A red boundary of such a hexagon indicates that GME was detected with $95\%$ confidence, and a blue boundary indicates that GME was not detected with confidence. When QREM is applied, $8$ heavy-hex unit cells are detected as GME.}
\label{fig:brisbane}
\end{figure}

\begin{figure}[htbp]
\centering
\subfigure[\texttt{ibm\_sherbrooke} (no QREM)]{
\label{fig:sherbrooke_no_qrem}
\includegraphics[scale=0.55]{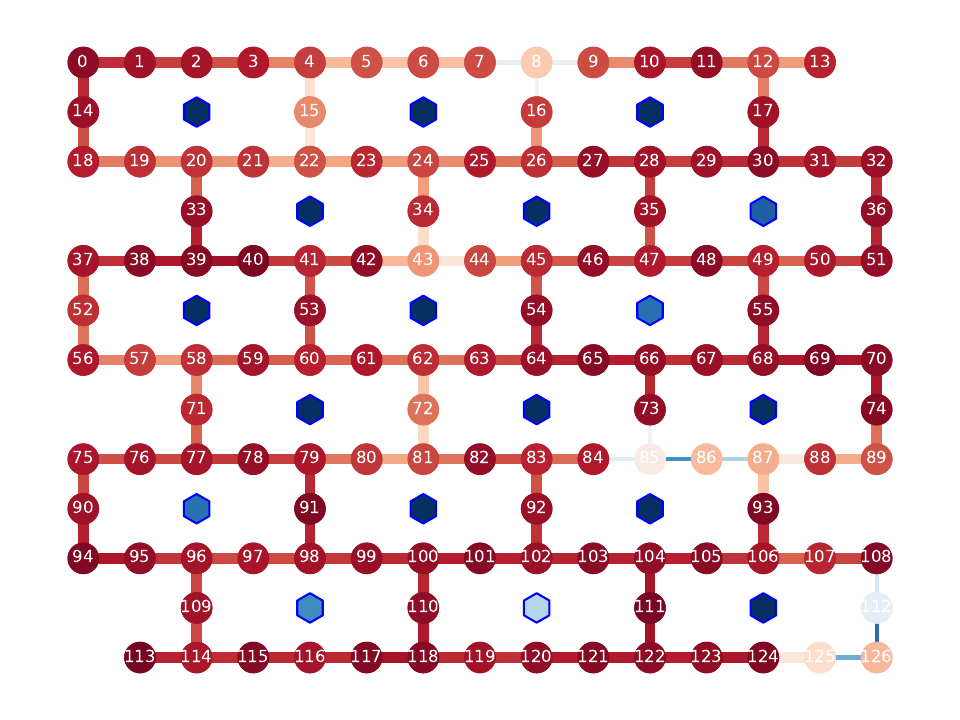}
}
\subfigure[\texttt{ibm\_sherbrooke} (QREM)]{
\label{fig:sherbrooke_qrem}
\includegraphics[scale=0.55]{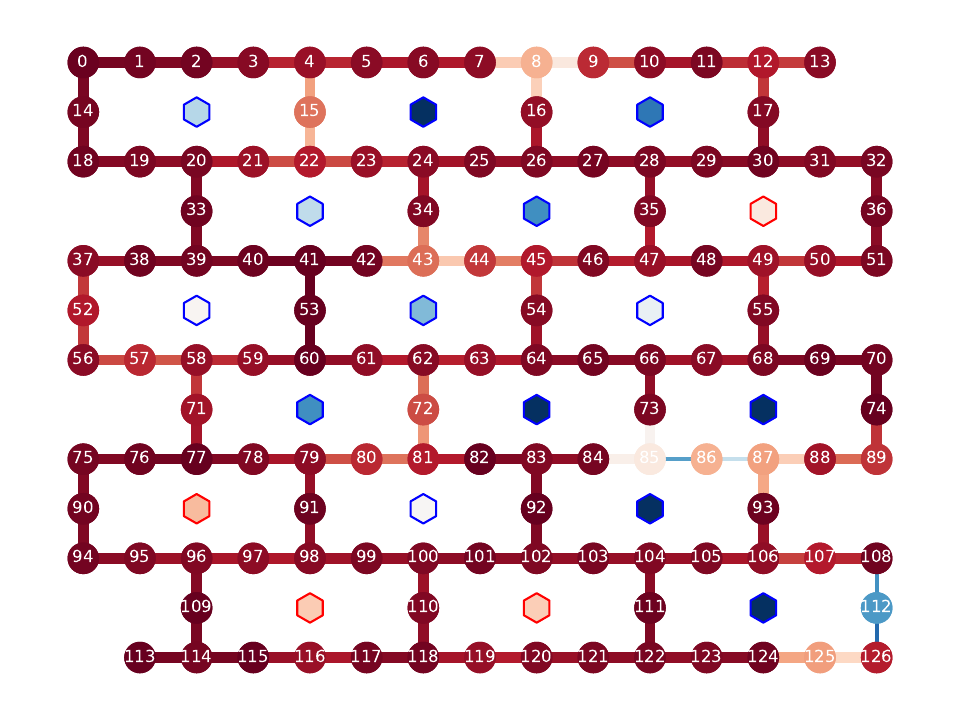}
}
\subfigure{
\includegraphics[scale=0.5]{images/color_scale.pdf}
}
\caption{A visualization of entanglement in the native graph state on the $127$-qubit \texttt{ibm\_sherbrooke} device. The largest bipartite entangled regions consist of (a) $122$ qubits, and (b) $125$ qubits. When QREM is applied, $4$ heavy-hex unit cells are detected as GME.}
\label{fig:sherbrooke}
\end{figure}

\begin{figure}[htbp]
\centering
\subfigure[\texttt{ibm\_cusco} (no QREM)]{
\label{fig:cusco_no_qrem}
\includegraphics[scale=0.55]{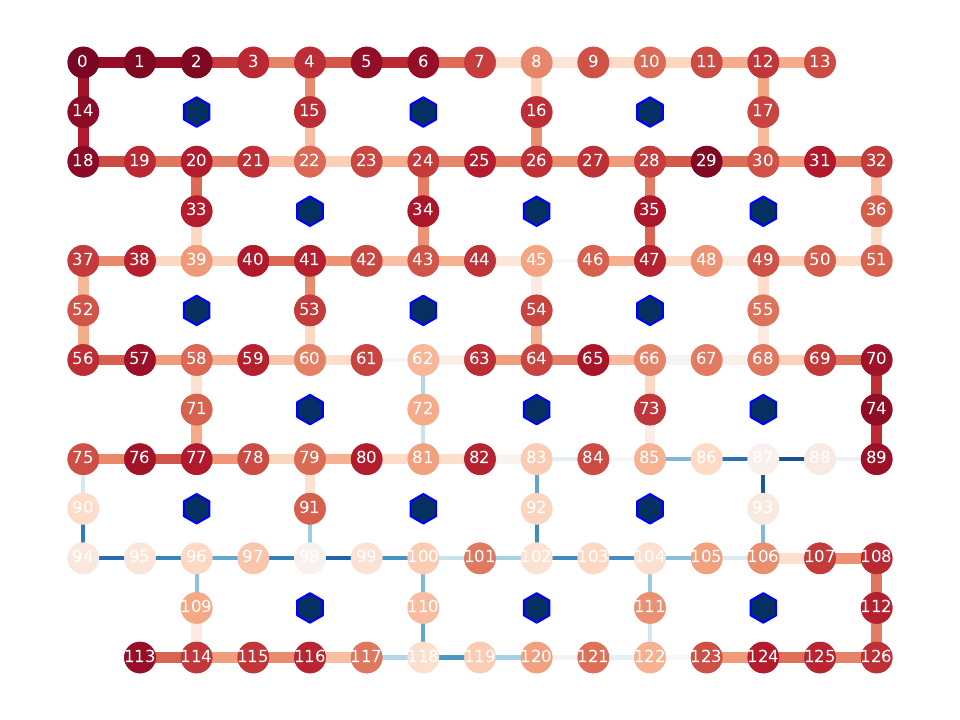}
}
\subfigure[\texttt{ibm\_cusco} (QREM)]{
\label{fig:cusco_qrem}
\includegraphics[scale=0.55]{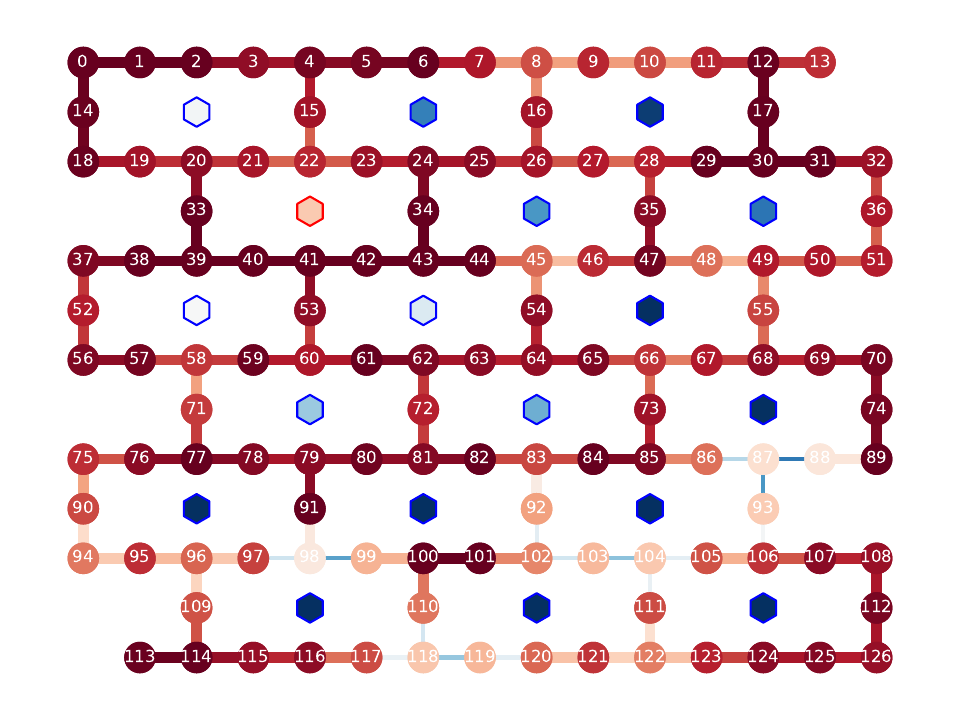}
}
\subfigure{
\includegraphics[scale=0.5]{images/color_scale.pdf}
}
\caption{A visualization of entanglement in the native graph state on the $127$-qubit \texttt{ibm\_cusco} device. The largest bipartite entangled regions consist of (a) $87$ qubits, and (b) $103$ qubits. When QREM is applied, $1$ heavy-hex unit cell is detected as GME.}
\label{fig:cusco}
\end{figure}

The results are illustrated in Figures \ref{fig:brisbane}, \ref{fig:sherbrooke} and \ref{fig:cusco}.
For the devices \texttt{ibm\_brisbane}, \texttt{ibm\_sherbrooke} and \texttt{ibm\_cusco}, the results indicate that $8$, $4$ and $1$ heavy-hex unit cells can be verified as genuinely multipartite entangled when QREM is applied, respectively.

Notably, the size of the largest $1$-D cluster state that can be verified as GME is similar for all three devices. 
In contrast, there is a remarkable difference in the number of heavy-hex unit cells that can be verified as GME, e.g., $8$ for \texttt{ibm\_brisbane} and $1$ for \texttt{ibm\_cusco}. This shows that the ability to generate multipartite entangled states is spread more evenly across the device for \texttt{ibm\_brisbane}. 
Therefore, for applications that require a larger number of qubits one would expect that \texttt{ibm\_brisbane} yields better results. 
In general, evaluating multipartite entanglement witnesses for different types of subgraphs can lead to more expressive results that can be interpreted in the context of practical applications.
For example, for simulations of a Heisenberg model on a $1$-D lattice, the size of the largest $1$-D cluster state verified as GME could be a suitable metric. When considering, e.g., a Kagome lattice, the size of the largest heavy-hex subgraph verified as GME could be a suitable metric.

\smallskip

(ii) The coloring-based witness (CBW)
\begin{equation}
\label{eq:experiments_W_col}
W^{c}(G',G)=\frac{3}{2}\I-P(V_1',G)-P(V_2',G)
\end{equation}
where $\V'=\{V_1',V_2'\}$ is the partition of the set of vertices $V'$ of $G'$ induced by the coloring of the graph.
This witness has a higher white noise tolerance $p_{\tol}> 1/4$. 
However, as the operator \ref{eq:experiments_W_col} cannot be decomposed as a sum of $m$-local observables for a fixed $m$ independent of the number of qubits $n'$, its expectation cannot be efficiently computed with the QREM described in Section \ref{sec:qrem}.
This can be remedied by considering a refinement of \ref{eq:experiments_W_col}, that is, the operator $W(\widetilde{V}',G)$ \ref{eq:witness_gs_gen} for a refinement $\widetilde{\V}'$ of the partition $\V'$. 
For $1$-D cluster states, the refinement is chosen by subdividing the state in groups of $5$ connected qubits and $\leq 5$ qubits in the remaining group. This construction is ambiguous: depending on the order of the qubits it yields two different refinements of the CBW. Therefore, we choose the minimum of both evaluated witnesses.
Compared to the SSW such a refinement of the CBW still has a higher white noise tolerance, e.g., $p_{\tol}=1.54/n'$ if $n'$ is a multiple of $5$ (Appendix \ref{sec:appendix_white_noise}).

Here, the CBW is utilized as follows.
We evaluate the CBW for the subgraphs isomorphic to $\Cl_n$, for $n=2,\dotsc,30$, that minimize the SSW without QREM.
Furthermore, we evaluate the refinement of the CBW for the subgraphs isomorphic to $\Cl_n$, for $n=2,\dotsc,30$, that minimize the SSW with QREM. 

The results are illustrated in Figure \ref{fig:cluster}.
For the devices \texttt{ibm\_brisbane}, \texttt{ibm\_sherbrooke} and \texttt{ibm\_cusco}, the results indicate that a $9$-qubit, $11$-qubit and $9$-qubit $1$-D cluster state can be verified as genuinely multipartite entangled without QREM, respectively. This is comparable to the results for the SSW. Notably, the difference between the expectations of the SSW and CBW increases with the number of qubits.

When QREM is applied, the results indicate that a $25$-qubit, $27$-qubit and $27$-qubit $1$-D cluster state can be verified as genuinely multipartite entangled for \texttt{ibm\_brisbane}, \texttt{ibm\_sherbrooke} and \texttt{ibm\_cusco}, respectively.

\smallskip

For \texttt{ibm\_brisbane}, the expectations of the SSW and the refinement of the CBW are almost identical independent of the number of qubits, despite its higher white noise tolerance. This indicates that the white noise tolerance is not a sufficient metric to assess the robustness of an entanglement witness under realistic experimental conditions. An avenue for future research could be the investigation of the robustness of entanglement witnesses under more realistic noise models.
Also note that in some cases the expectation of the CBW is larger than the expectation of the SSW. On first sight, this seems contradictory to the fact that $W^{\stab}(G',G)\geq W^{c}(G',G)$, as the SSW is a refinement of the CBW. Yet, this translates to a similar relation for the expectations, i.e., $\braket{W^{\stab}(G',G)}\geq\braket{W^{c}(G',G)}$, only if they are evaluated for probability distributions.
With QREM we obtain quasiprobability distributions that may include negative probabilities.

\smallskip

For \texttt{ibm\_sherbrooke} and \texttt{ibm\_cusco}, the expectations of the SSW and CBW are comparable, yet, the expectations of the CBW are consistently lower. 
At this point, it is not sufficiently investigated if this is a reliable result.
This difference could very well be attributed to the QREM method:
for the evaluation of witnesses we cap expectations of projectors $P(U)=P(U,G)$ at $1$. For the SSW, the projectors are evaluated for each qubit separately. For the refinement of the CBW, the projectors correspond to subsets of $2$ or $3$ qubits. 
Then, for example, if one considers a projector for a subset of $2$ qubits $P(\{i,j\})$ with $\braket{P(\{i\})}>1$ (before capping) and $\braket{P(\{j\})}<1$ , it may occur that $\braket{P(\{i,j\})}\geq1$. After capping we have $\braket{P(\{i\})}=1$, 
$\braket{P(\{j\})}<1$ and $\braket{P(\{i,j\})}=1$, so that in this case the CBW is lower than the SSW. 
This phenomenon should be further investigated, e.g, by comparing the findings for different readout error mitigation methods.


\subsection{Benchmarking}
\label{sec:benchmarking}

In the following, the experiments are considered with regard to various aspects of benchmarking \cite{Colin2022} such as scalability, verifiability and comparability.

\smallskip

\subsubsection{Architecture-specific benchmarks}
\begin{figure}
\centering
\subfigure{
\includegraphics[width = 0.55\columnwidth]{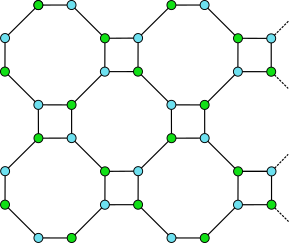}
}
\subfigure{
\includegraphics[width = 0.55\columnwidth]{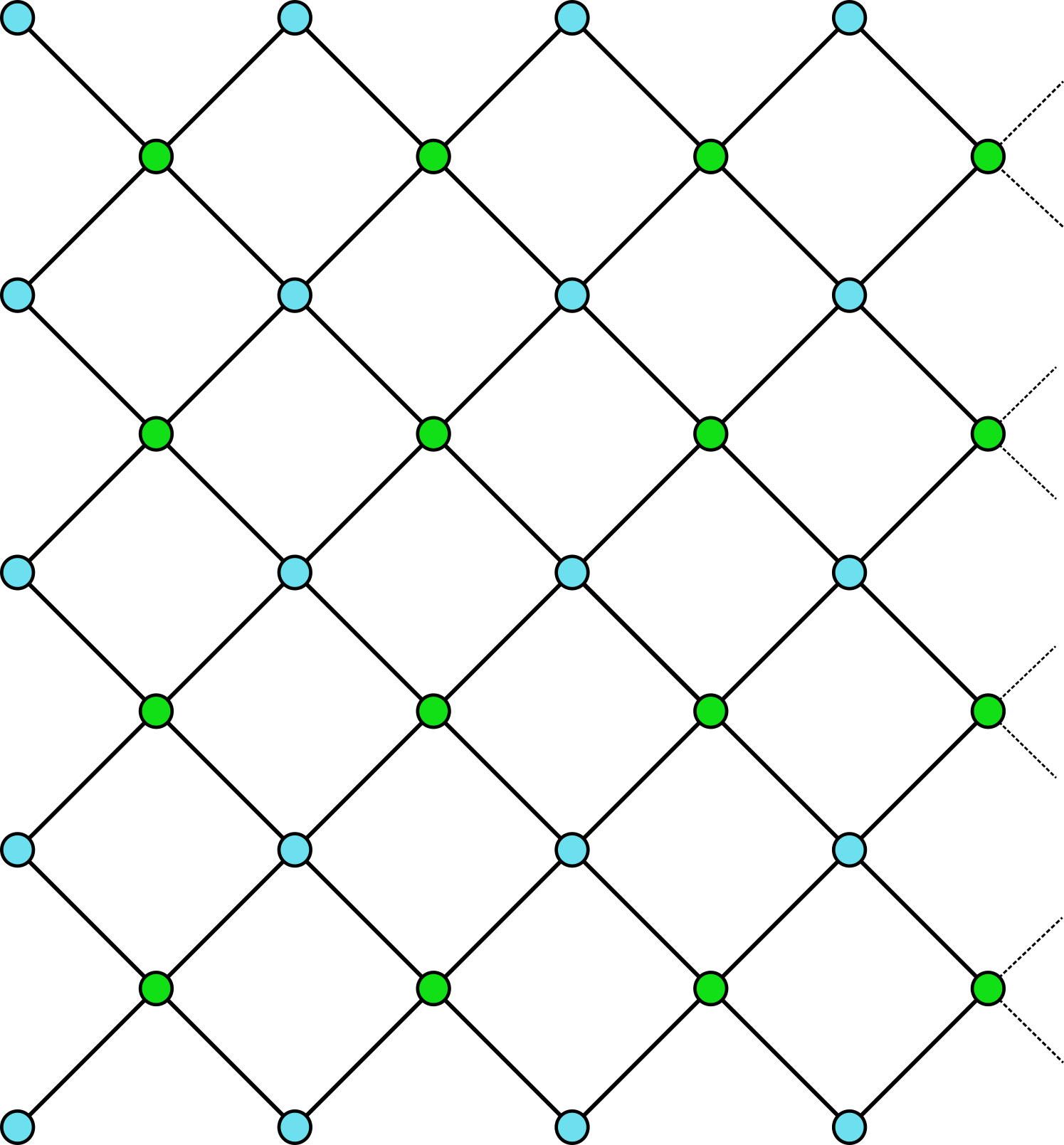}
}
\caption{The coupling map for the \textit{Rigetti Aspen} (top) and \textit{Google Sycamore} (bottom) device is $2$-colorable.}
\label{fig:nisq_topologies}
\end{figure}

By performing benchmarks with graph states that correspond to the native qubit topology of the QPU under test, computational overhead for classical preprocessing with circuit optimization and qubit routing is reduced without introducing additional SWAP gates. Furthermore, the execution of CZ gates can be straightforwardly parallelized with respect to the qubit topology as is shown in Figure \ref{fig:design}. For $2$-colorable graphs, only two measurement settings - independent of the number of qubits - are needed. Prominent examples apart from IBM Quantum devices that have $2$-colorable coupling graphs are shown in Figure \ref{fig:nisq_topologies}. Note that especially for ion trap based QPUs, all-to-all connectivity can be achieved in the NISQ era. With this, every graph state can be natively implemented without introducing additional SWAP gates.
 
The whole QPU can be benchmarked for A) bipartite entanglement so that regions of connected qubits that are bipartite entangled are found. Ideally, all benchmarked qubits are bipartite entangled such as shown in \ref{fig:brisbane_qrem}. With the same measurement results, the QPU can be benchmarked for B) genuine multipartite entanglement so that regions of connected qubits that are genuinely multipartite entangled are found. That is, the graph state induced by such a subset of qubits can be prepared on the QPU and verified as GME. Realistically, for NISQ devices, these subsets correspond to smaller subgraphs such as shown for the heavy-hex unit cells in Figure \ref{fig:brisbane_qrem} and the subgraphs isomorphic to $1$-D cluster states in Figure \ref{fig:cluster}. Based on our observations, we advise using the stabilizer sum witness for performing the benchmarks as it can be evaluated efficiently in a scalable manner also with readout error mitigation. The coloring-based witness is more costly to evaluate (especially with readout error mitigation) and has not shown a significant advantage in detecting GME.

With our method, the capability of generating entangled states based on natively implementable graph states can be assessed and compared to the results from different suitable architectures. If the comparison is done between hardware platforms where one platform can only implement the graph state by using SWAP operations, the comparison is not straightforward anymore. If the results of said benchmarks would be worse on this platform, it is not clear if this can only be explained with the CNOT gate overhead introduced by the additional SWAP gates, or if the device would also perform worse independent of this gate overhead. Only if such a device performs better despite an additional SWAP overhead, the results can be interpreted comparatively with other devices in the sense that it performs better in said entanglement generation tasks. Hence, we advise using this as an architecture-specific benchmark.
 
\smallskip
 
\subsubsection{Architecture-independent benchmarks}
Multipartite entanglement generation for 1-D cluster states can be benchmarked on every hardware topology, hence generating the longest chain of qubits that exhibits GME can be seen as an architecture-independent benchmark. The context is important here: if entanglement for a $1$-D cluster state is verified as part of a larger experiment that probes an overarching graph state, then these results are not necessarily comparable to just verifying entanglement for such a cluster state, mainly due to increased (non-local) noise from the additional gate executions surrounding this subgraph.

\section{Outlook}
\label{sec:Outlook}
	
In summary, we discussed a scalable method for benchmarking the entanglement generation capabilities of NISQ devices using entanglement witnesses. This method was tested on different IBM QPUs for analyzing bipartite entanglement over all qubits and the ability to generate GME for $1$-D cluster states and heavy-hex unit cells. In addition, we discussed the implications of using this method as a benchmark.
Finally, based on the results presented, we list potential further approaches that can be pursued in future research endeavors.

\begin{enumerate}
\item \underline{Benchmarking different QPUs}: Since the developed method can be executed efficiently on several NISQ devices, performing additional benchmarks on these QPUs will provide insightful data. Based on this, comparisons between different devices with the criteria discussed in Section \ref{sec:benchmarking} can be drawn.

\item \underline{Maintaining Entanglement}: Measuring the duration for which verified bipartite entanglement for a graph state or GME for a subet of qubits can be maintained on a QPU under test is an interesting extension for the proposed benchmark. For this, the experiments can be augmented by delayed measurements with an incremental increase in delay time in order to obtain time-dependent data. Similar experiments were performed based on different entanglement verification criteria \cite{KKHMH23} and can be compared with the presented method.

\item \underline{Parallel Circuit Execution}: The verification of entanglement in specific subgraphs could potentially be used for the evaluation of parallelization possibilities on a QPU. The simultaneous execution of multiple spatially separated quantum circuits on a single QPU (often called multi programming) is an active field of research and several compilers that perform such parallel scheduling tasks were proposed such as \textit{palloq} \cite{Ohkura2022}, \textit{QuMC} \cite{Niu2023} and \textit{QuCloud/QuCloud+} \cite{Liu2022}. The analysis of regions on a QPU that show good entanglement generation capabilities could be used for more efficient scheduling implementations. A possible choice for such regions are given by the heavy-hex unit cells that can be verified as GME in Section \ref{sec:experiments}.
\end{enumerate}

\section*{Acknowledgments}

This work was funded by the Federal Ministry for Economic Affairs and Climate Action (German: Bundesministerium für Wirtschaft und Klimaschutz) under the project funding number 01MQ22007A. The authors are responsible for the content of this publication.

\section{Appendix}
\label{sec:Appendix}

\renewcommand{\stab}{\Sigma}
\newcommand{\pos}{\pi}
\newcommand{\sort}{\mathrm{sort}}
\newcommand{\marginal}{\mathrm{marginal}}
\newcommand{\reduce}{\mathrm{reduce}}
\newcommand{\mitigate}{\mathrm{mitigate}}
\newcommand{\subsetwitness}{\mathrm{subset\_witness}}
\newcommand{\U}{\mathcal U}

\newcommand{\A}{\mathcal A}
\newcommand{\B}{\mathcal B}


\subsection{Evaluation of entanglement witnesses}
\label{sec:implementation}

Subsequently, we discuss the main aspects of implementing the evaluation of entanglement witnesses and readout error mitigation \ref{sec:qrem}.

In our setting, we consider a graph $G=(V,E)$ and a partition $\V=\{V_1,\dotsc,V_k\}$ of the set of vertices $V$ corresponding to a vertex coloring of $G$ with $k$ colors (here, $k=2$). 
The measurement results for the prepared state $\rho$ with respect to the graph state $\ket{G}$ are given by a set of probability distributions $\Delta=\{\Delta_1,\dotsc,\Delta_k\}$. Each probability distribution $\Delta_l=\{(x,p(x))\mid x\in\{0,1\}^n, p(x)\neq0\}$ contains measurement results in the measurement setting $\otimes_{i\in V_l}\x_i\otimes_{j\in V\setminus V_l}\z_j$. For readout error mitigation we utilize the calibration matrices $(A^{(i)})_{i\in V}$.
Then entanglement witnesses are evaluated as follows.

\smallskip
\subsubsection{Coloring-based witness}

Algorithm \ref{alg:witness} computes the coloring-based witness for a subgraph $G'=(U,E')$. More specifically,
let $\U=\{U_1,\dotsc,U_k\}$ be the partition of $U$ induced by the coloring of the graph $G$, i.e., we have $U_l=V_l\cap U$.
Then we compute the expectation of the operator
\begin{equation}
\label{eq:witness_appendix}
W(\U,G)=\left(k-\frac12\right)\I-\sum\limits_{l=1}^kP(U_l,G).
\end{equation}
Note that each stabilizer projector $P(U_l,G)$ acts non-trivially on all qubits in $U_l$ and their neighbors $N(U_l)$ in the graph $G$. Then the QREM as described in Section \ref{sec:qrem} scales exponentially in the number of these qubits.
Hence, it can only be applied for small subsets of qubits. For larger subsets of qubits, one can compute a (refinement of the) coloring-based witness with QREM by subdividing the set of qubits $U$ into smaller subsets and computing the witnesses for each subset. For example, for a subdivision $U=A\cup B$ consider the partitions $\A=\{U_1\cap A,\dotsc,U_l\cap A\}$ and $\B=\{U_1\cap B,\dotsc,U_l\cap B\}$ of the sets $A$ and $B$, respectively.
Then the partition $\A\cup\B$ is a refinement of the partition $\U$ of the set $U=A\cup B$. With equations \ref{eq:witness_refinement} and \ref{eq:witness_partition}, we find
\begin{equation}
W(\U,G)\leq W(\A\cup\B,G)=\I/2+W(\A,G)+W(\B,G).
\end{equation} 
The witnesses $W(\A,G)$ and $W(\B,G)$ act non-trivially on a smaller number of qubits.

\begin{figure}
\begin{algorithm}[H]
\caption{\textcolor{blue}{$\mathrm{witness}$}}
\begin{algorithmic}
\State \textbf{Input:}
A graph $G=(V,E)$, a partition $\V$ of $V$, distributions $\Delta_{\V}$, a subset of qubits $U\subset V$, calibration matrices $(A^{(i)})_{i\in V}$, and a Boolean $qrem$.
\State \textbf{Output:} 
Expectation of witness $W=\braket{W(\U,G)}$.
\State $\Delta=\Delta_{\V}$
\State $\stab=\{\}$ \Comment{Stabilizers}
\For{$j$ in $U$}
	\State $\stab[j]=S_j$
\EndFor
\State $\U=\{\}$ \Comment{Partition $\U=\{U_1,\dotsc,U_k\}$}
\For{$l=1$ to $k$} 	
	\State $\U[l]=V_l\cap U$
\EndFor
\If{$qrem$} \Comment{Readout error mitigation}
    \State $\Delta, \stab = \textcolor{blue}{\mathrm{qrem}}(G,\Delta_{\V},\stab,(A^{(i)})_{i\in V},\U)$
\EndIf
\State $W=k-\frac12$ 
\For{$l=1$ to $k$} 
	\If{$\U[l]\neq\emptyset$}
		\State $P=0$ \Comment{Evaluate projector $P(U_l)$}
		\For{$(x,p(x))$ in $\Delta[l]$}
			\State $temp=1$ \Comment{Evaluate $\braket{x|P(U_l)|x}$}
			\For {$i$ in $\U[l]$}	 
				\State $temp=temp\cdot\dfrac{1+\braket{x|\stab[i]|x}}{2}$
			\EndFor
			\State $P=P+p(x)\cdot temp$
		\EndFor
		\State $W=W-\min(P,1)$ \Comment{Cap projector at $1$}
	\Else
		\State $W=W-1$ \Comment{$P(\emptyset)=1$}
	\EndIf
\EndFor
\State \textbf{return:} $W$
\end{algorithmic}
\label{alg:witness}
\end{algorithm}
\caption*{This algorithm computes the expectation of the coloring-based witness \ref{eq:witness_appendix} for a subset $U$ of qubits, and the expectation of the stabilizer $S_i$ if $U=\{i\}$, for $i\in V$. For this, we compute the expectation of the stabilizer projectors for each color.
The expectation of the projector $P(U_l)=P(U_l,G)$ is calculated from the distribution $\Delta_l$ of measurement outcomes in measurement setting $\oplus_{i\in V_l}\oplus_{j\in V\setminus V_l}Z_j$. This corresponds to a change of basis such that in this basis the stabilizers $S_i$, for $i\in V_l$, are a product of $\I$ and Pauli-$Z$ operators. Thus, the projector $P(U_l)$ is a product of diagonal operators. Therefore, calculating its expectation on a computational basis state $\ket{x}$ is accomplished by taking the product of the expectations of the diagonal operators $(\I+S_i)/2$.
}
\end{figure}

\smallskip
\subsubsection{Stabilizer sum witness}

If the set $U$ consists of exactly one qubit, i.e., $U=\{i\}$, for $i\in V$,
equation \ref{eq:witness_appendix} simplifies to 
\begin{equation}
W(\U,G)=\frac12-P(\{i\},G)=-\frac{S_i}{2}.
\end{equation}
 Here, we use that $P(\emptyset,G)=1$.
Thus, we can apply Algorithm \ref{alg:witness} to compute the expectations of all stabilizers $S_i$, for $i\in V$. In our experiments, each such stabilizer acts non-trivially only on at most $4$ qubits since the maximum vertex degree of a heavy-hex graph is $3$. Therefore, readout error mitigation as described in Section \ref{sec:qrem} can be utilized. 
The (mitigated) expectations of the stabilizers can further be used to calculate stabilizer sum witnesses for subgraphs.

\smallskip
\subsubsection{QREM}

The quantum readout error mitigation described in Section \ref{sec:qrem} is implemented as shown in Algorithm \ref{alg:qrem}. 
For this, we assume that the following functions are given:

\begin{itemize}

\item \textcolor{blue}{$\sort$}: \textbf{Input:} set of integers. \textbf{Output:} sorted list of integers.

\item \textcolor{blue}{$\reduce$}: \textbf{Input:} stabilizer, (sorted) list of positions. \textbf{Output:} reduced stabilizers with respect to the given positions. For example, for a stabilizer $\z_7\x_8\z_9$ and positions $\pos=(7,8,9)$, the reduced stabilizer is $\z_0\x_1\z_2$.

\item \textcolor{blue}{$\marginal$}: \textbf{Input:} distribution, (sorted) list of positions. \textbf{Output:} marginal distributions with respect to the given positions. 

\item \textcolor{blue}{$\mitigate$}: \textbf{Input:} distribution, list of calibration matrices. \textbf{Output:} mitigated distribution.

\end{itemize}

\begin{figure}
\begin{algorithm}[H]
\caption{\textcolor{blue}{$\mathrm{qrem}$}}
\begin{algorithmic}
\State \textbf{Input:} A graph $G$, distributions $\Delta_{\V}$, stabilizers $\stab$, calibration matrices $(A^{(i)})_{i\in V}$, and a partition $\U$ of $U$.
\State \textbf{Output:} 
Mitigated distributions $\Delta_{\U}$ and reduced stabilizers $\stab_{\U}$.
\State $\Delta_{\U}=\{\}$
\State $\stab_{\U}=\{\}$
\For{$l=1$ to $k$}
	\If{$\U[l]\neq\emptyset$}
		\State $\pos=\sort(U_l\cup N(U_l))$
		\For{$i$ in $U_l$} \Comment{Reduced stabilizers}
			\State $\stab_{\U}[i]=\reduce(\stab[i],\pos)$
		\EndFor
		\State $\Delta_{\U}[l]=\marginal(\Delta_{\V}[l],\pos)$
		\State $\Delta_{\U}[l]=\mitigate(\Delta_{\U}[l],(A^{(\pos_1)},\dotsc,A^{(\pos_k)}))$
	\EndIf
\EndFor
\State \textbf{return:} $\Delta_{\U}$, $\stab_{\U}$
\end{algorithmic}
\label{alg:qrem}
\end{algorithm}
\caption*{This algorithm computes the reduced stabilizers and the mitigated (marginal) distribution with respect to a subset $U$ of qubits.}
\end{figure}

\subsection{White noise tolerance}
\label{sec:appendix_white_noise}

From the definition of the white noise tolerance for a graph state $\ket{G}$ and a witness $W$ we find 
\begin{equation}
p_{\tol}=\left(1-\frac{\tr(W)}{2^n\tr(W\ket{G}\bra{G})}\right)^{-1}.
\end{equation}
For a witness $W$ of the form \ref{eq:witness} we have $\tr(W\ket{G}\bra{G})=-1/2$. 
It remains to calculate
\begin{equation}
\tr(W)=2^n\left(k-\frac12\right)-2^n\sum\limits_{l=1}^k2^{-n_l}
\end{equation}
where $n_l=|V_l|$ is the number of qubits in each vertex set of the partition $\V$, and we use that $\tr(P(V_l,G))=2^{n-n_l}$.
Then we obtain
\begin{equation}
p_{\tol}=\frac12\left(k-\sum\limits_{l=1}^k2^{-n_l}\right)^{-1}>\frac{1}{2k}.
\end{equation}
In particular, for the stabilizer sum witness, i.e., $\V=\{\{i\}\}_{i\in V}$, we find $p_{\tol}=1/n$.

In Section \ref{sec:experiments_results}, we consider a refinement of the coloring-based witness for $1$-D cluster states. This refinement is obtained by subdividing a state in groups of $5$ connected qubits and $\leq 5$ qubits in the remaining group. This corresponds to a partition $\V=\{V_0,\dotsc,V_{\lceil n/5\rceil-1}\}$, where $V_{2i}, V_{2i+1}$, for $i=0,\dotsc,\lceil n/5\rceil-1$, are the sets of qubits of the $i$-th group for each color.
Then we have $k=2\lceil n/5\rceil$. If $n$ is a multiple of $5$, we can assume that $|V_{2i}|=2$ and $|V_{2i+1}|=3$.
In this case, we have
\begin{equation}
p_{\tol}=\frac12\left(\frac{2n}{5}-\frac{n}{5}\left(\frac{1}{4}+\frac{1}{8}\right)\right)^{-1}=\frac{20}{13n}\approx 1.54n^{-1}.
\end{equation}
The values of $c(n)=n\cdot p_{\tol}$, for $n=1,\dotsc,24$, are shown in Table \ref{tab:ptol}.

\begin{table}[ht]
\centering
\renewcommand{\arraystretch}{1.5}
\begin{tabular}{|c|c|c|c|c|c|c|c|c|}
\hline 
$n$&$1$&$2$&$3$&$4$&$5$&$6$&$7$&$8$\\
\hline 
$c(n)$&$1.0$&$1.0$&$1.2$&$1.33$&$1.54$&$1.41$&$1.33$&$1.39$\\ 
\hline
$n$&$9$&$10$&$11$&$12$&$13$&$14$&$15$&$16$\\
\hline 
$c(n)$&$1.44$&$1.54$&$1.47$&$1.41$&$1.44$&$1.47$&$1.54$&$1.49$\\
\hline
$n$&$17$&$18$&$19$&$20$&$21$&$22$&$23$&$24$\\
\hline 
$c(n)$&$1.45$&$1.47$&$1.49$&$1.54$&$1.5$&$1.47$&$1.48$&$1.5$\\
\hline 
$n$&$25$&$26$&$27$&$28$&$29$&$30$&&\\
\hline 
$c(n)$&$1.54$&$1.51$&$1.48$&$1.49$&$1.51$&$1.54$&&\\
\hline 
\end{tabular} 
\caption{The factors $c(n)=n\cdot p_{\tol}$, for $n=1,\dotsc 30$.}
\label{tab:ptol}
\end{table}

\subsection{Properties of projectors}

For Hermitian operators $A,B$ on a finite-dimensional Hilbert space $\H$, we use the notation $A\geq B$ indicating that
$(A-B)$ is positive semidefinite.
The following result was shown in \cite{ZZYM19} (Proof of Proposition 2). 

\begin{lemma}
\label{lem:operators}
Let $P_1,\dotsc,P_k$ be commuting Hermitian operators on a finite-dimensional Hilbert space $\H$ with all eigenvalues in $\{0,1\}$.
Then we have
\begin{equation}
\prod\limits_{l=1}^kP_l\geq\sum\limits_{l=1}^kP_l-(k-1)\I.
\end{equation}
\end{lemma}

\subsection{A necessary condition for separability}

We prove a necessary condition for separability that can be used to construct entanglement witnesses for bipartite entanglement.

\begin{rk}
\label{rk:density}
We write $\sigma_i$ for $i=0,1,2,3$, where $\sigma_0=\I$ is the identity, and $\sigma_1=\sigma_x$, $\sigma_2=\sigma_y$, $\sigma_3=\sigma_z$ are the three Pauli matrices.
Consider the Hilbert space $\H=(\C^2)^{\otimes n}$. We write $\i=(i_1,\dotsc,i_n)$ for a multi-index.
The set of matrices $E_{\i}=\sigma_{i_1}\otimes\dotsb\otimes\sigma_{i_n}$ is orthogonal with respect to the Hilbert-Schmidt inner product,
that is, $(E_{\i},E_{\j})=\tr(E_{\i}^{\dag}E_{\j})=2^n\delta_{\i\j}$, and forms a basis of the real vector space of Hermitian matrices in $\H$.
Then any density operator may be represented as
\begin{equation}
\label{eq:density}
\rho=\frac{1}{2^n}\left(\I+\sum\limits_{\i\neq 0}\lambda_{\i}\sigma_{i_1}\otimes\dotsb\otimes\sigma_{i_d}\right)
\end{equation}
where $\lambda_{\i}$ are real numbers. Equation \ref{eq:density} does not include the non-negativity condition.
\end{rk}

\begin{prop}
\label{prop:separability}
Let $S, S'$ be Pauli product operators of the form:
\begin{equation}
S=\prod\limits_{m\in M}\sigma(m),\qquad
S'=\prod\limits_{m\in M}\sigma'(m)
\end{equation}
where $\sigma(m), \sigma'(m)\in\{\I,\sigma_x,\sigma_y,\sigma_z\}$ are Pauli operators acting on qubit $m$.
Let $A, B$ be a partition of M. Suppose that the state $\rho\in (\C^2)^{\otimes |M|}$ is separable with respect to $A, B$, that is, $\rho=\sum_kp_k\rho_k^A\otimes\rho_k^B$.
Consider the Pauli product operators
\begin{equation}
S_K=\prod\limits_{m\in M\cap K}\sigma(m),\quad
S_K'=\prod\limits_{m\in M\cap K}\sigma'(m),
\end{equation}
for $K\in\{A,B\}$. The operators $S_K$ and $S_K'$ are obtained from $S$ and $S'$, respectively, by replacing all Pauli operators acting on qubits in $M\setminus K$ to identities.
If the anti-commutation relations 
\begin{equation}
\label{eq:anti-commutation}
\{S_K,S_K'\}=0,\quad \text{for } K\in\{A,B\},
\end{equation}
are satisfied, then we have
\begin{equation}
\langle S\rangle + \langle S'\rangle\leq 1.
\end{equation}

\begin{proof}
We consider the case $\rho=\rho^A\otimes\rho^B$.
Then we have
\begin{equation}
\label{eq:ST}
\langle S\rangle + \langle S'\rangle
=\langle S_A\rangle \langle S_B\rangle + \langle S_A'\rangle \langle S_B'\rangle.
\end{equation}
Define the Hermitian operators 
\begin{equation}
O_K(\theta)=\cos(\theta)S_K+\sin(\theta)S_K',
\end{equation}
for $K\in\{A,B\}$. The operators $O_K(\theta)$ have all eigenvalues in $\{-1,1\}$:
we have 
\begin{align}
\begin{split}
(O_K(\theta))^2=&(\cos(\theta))^2S_K^2+\cos(\theta)\sin(\theta)(S_KS_K'+S_K'S_K)\\
&+(\sin(\theta))^2S_K'^2=\I.
\end{split}
\end{align}

The Pauli products $S_K, S_K'$ are of the form $E_{\i}=\sigma_{i_1}\otimes\dotsb\otimes\sigma_{i_n}$, for some indexes $\i=\i_K, \i_K'$.
Then with \ref{eq:density} we may write
\begin{equation}
\rho^K=\frac{1}{2^{|K|}}\left(\I+r_KO_K(\theta_K)+\smashoperator{\sum\limits_{\i\neq0,\i_K,\i_K'}}\lambda_{\i}^K\sigma_{i_0}\otimes\dotsb\otimes\sigma_{i_{|K|}}\right),
\end{equation}
for real numbers $\lambda_{\i}^K$ and $r_K\geq0$, $\theta_K\in [0,2\pi)$.
Then with \ref{eq:ST} we find
\begin{align}
\begin{split}
\langle S\rangle + \langle S'\rangle&=r_Ar_B(\cos(\theta_A)\cos(\theta_B)+\sin(\theta_A)\sin(\theta_B))\\
&=r_Ar_B\cos(\theta_A-\theta_B)\leq r_Ar_B.
\end{split}
\end{align}
In the following we show that $r_K\leq 1$ for $K\in\{A,B\}$: on the one hand we have 
$\tr(\rho^KO_K(\theta_K))=r_K(\cos^2(\theta_K)+\sin^2(\theta_K))=r_K$. 
On the other hand the Hermitian operator $O_K(\theta_K)$ has a spectral decomposition $\sum_l\lambda_l\ket{v_l}\bra{v_l}$, where $\lambda_l$ are the real eigenvalues of $O_K(\theta_K)$ and the vectors $\ket{v_l}$ form an orthonormal basis. Then we have:
\begin{equation}
\tr(\rho^KO_K(\theta_K))=\sum\limits_l\lambda_l\tr(\rho^K\ket{v_l}\bra{v_l})\leq\max\limits_l\lambda_l=1.
\end{equation}
Here, we used that the density operator $\rho^K$ is positive and has trace equal to one.
This finishes the proof for the case $\rho=\rho^A\otimes\rho^B$. 
Then the claim follows from the linearity of expectations.
\end{proof}
\end{prop}

As a special case, we obtain the following necessary condition for separability in the context of graph states.
This is a generalization of a similar condition for full separability \cite{TG05,Hein06}.

\begin{prop}
\label{prop:separability_graph}
Let $G=(V,E)$ be a graph and $(i,j)\in E$.
Consider the stabilizer operators
\begin{equation}
S_i=\sigma_x^i\prod\limits_{l\in N_i}\sigma_z^l,\qquad
S_j=\sigma_x^j\prod\limits_{l\in N_j}\sigma_z^l.
\end{equation}
Let $A,B$ be a partition of $V$ with $i\in A$ and $j\in B$. Suppose that the state $\rho\in (\C^2)^{\otimes n}$ is separable with respect to $A,B$, that is, $\rho=\sum_kp_k\rho_k^A\otimes\rho_k^B$.
Then we have 
\begin{equation}
\langle S_i\rangle + \langle S_j\rangle\leq 1.
\end{equation}

\begin{proof}
With $S=S_i$ and $S'=S_j$ we find that
\begin{align}
S_A&=\sigma_x^i\prod\limits_{l\in N_i\cap A}\sigma_z^l,
&S_A'&=\sigma_z^i\prod\limits_{l\in N_j\setminus\{i\}\cap A}\sigma_z^l,\\
S_B&=\sigma_z^j\prod\limits_{l\in N_i\setminus\{j\}\cap B}\sigma_z^l,
&S_B'&=\sigma_x^j\prod\limits_{l\in N_j\cap B}\sigma_z^l.
\end{align}
Clearly, the anti-commutation relations \ref{eq:anti-commutation} are satisfied. Then the claim follows from Proposition \ref{prop:separability}.
\end{proof}
\end{prop}


\bibliographystyle{IEEEtran}
\bibliography{entanglement_paper.bib}

\begin{thebibliography}{10}
\providecommand{\url}[1]{#1}
\csname url@samestyle\endcsname
\providecommand{\newblock}{\relax}
\providecommand{\bibinfo}[2]{#2}
\providecommand{\BIBentrySTDinterwordspacing}{\spaceskip=0pt\relax}
\providecommand{\BIBentryALTinterwordstretchfactor}{4}
\providecommand{\BIBentryALTinterwordspacing}{\spaceskip=\fontdimen2\font plus
\BIBentryALTinterwordstretchfactor\fontdimen3\font minus
  \fontdimen4\font\relax}
\providecommand{\BIBforeignlanguage}[2]{{%
\expandafter\ifx\csname l@#1\endcsname\relax
\typeout{** WARNING: IEEEtran.bst: No hyphenation pattern has been}%
\typeout{** loaded for the language `#1'. Using the pattern for}%
\typeout{** the default language instead.}%
\else
\language=\csname l@#1\endcsname
\fi
#2}}
\providecommand{\BIBdecl}{\relax}
\BIBdecl

\bibitem{MWHH21ghz}
G.~J. Mooney, G.~A. White, C.~D. Hill \emph{et~al.},
  ``\href{https://dx.doi.org/10.1088/2399-6528/ac1df7}{Generation and
  verification of 27-qubit {Greenberger-Horne-Zeilinger} states in a
  superconducting quantum computer},'' \emph{Journal of Physics
  Communications}, vol.~5, no.~9, p. 095004, 2021.

\bibitem{MWHH21}
------,
  ``\href{https://onlinelibrary.wiley.com/doi/abs/10.1002/qute.202100061}{Whole-Device
  Entanglement in a 65-Qubit Superconducting Quantum Computer},''
  \emph{Advanced Quantum Technologies}, vol.~4, no.~10, p. 2100061, 2021.

\bibitem{CWC23}
S.~Cao, B.~Wu, F.~Chen \emph{et~al.},
  ``\href{https://www.nature.com/articles/s41586-023-06195-1}{Generation of
  genuine entanglement up to 51 superconducting qubits},'' \emph{Nature}, vol.
  619, pp. 738--742, 2023.

\bibitem{KKHMH23}
J.~F. Kam, H.~Kang, C.~D. Hill \emph{et~al.},
  ``\href{https://arxiv.org/abs/2312.15170}{Generation and preservation of
  large entangled states on physical quantum devices},'' \emph{arXiv preprint
  arXiv:2312.15170}, 2023.

\bibitem{Hamilton22}
K.~E. Hamilton, N.~Laanait, A.~Francis \emph{et~al.},
  ``\href{https://arxiv.org/abs/2209.00678}{An entanglement-based volumetric
  benchmark for near-term quantum hardware},'' \emph{arXiv preprint
  arXiv:2209.00678}, 2022.

\bibitem{VolumetricFramework}
R.~Blume-Kohout and K.~C. Young,
  ``\href{https://doi.org/10.22331/q-2020-11-15-362}{A volumetric framework for
  quantum computer benchmarks},'' \emph{{Quantum}}, vol.~4, p. 362, 2020.

\bibitem{OneWayQC}
R.~Raussendorf and H.~J. Briegel,
  ``\href{https://link.aps.org/doi/10.1103/PhysRevLett.86.5188}{A One-Way
  Quantum Computer},'' \emph{Physcal Review Letters}, vol.~86, pp. 5188--5191,
  2001.

\bibitem{Hein06}
M.~Hein, W.~D{\"u}r, J.~Eisert \emph{et~al.},
  ``\href{https://arxiv.org/abs/quant-ph/0602096}{Entanglement in graph states
  and its applications},'' \emph{arXiv preprint quant-ph/0602096}, 2006.

\bibitem{SupercheQ}
P.~Gokhale, E.~R. Anschuetz, C.~Campbell \emph{et~al.},
  ``\href{https://doi.org/10.48550/arXiv.2212.03850}{{SupercheQ}: Quantum
  Advantage for Distributed Databases},'' 2022.

\bibitem{TG05}
G.~T{\'o}th and O.~G{\"u}hne,
  ``\href{https://link.aps.org/doi/10.1103/PhysRevLett.94.060501}{Detecting
  genuine multipartite entanglement with two local measurements},''
  \emph{Physical Review Letters}, vol.~94, p. 060501, 2005.

\bibitem{B04}
M.~Bourennane, M.~Eibl, C.~Kurtsiefer \emph{et~al.},
  ``\href{https://link.aps.org/doi/10.1103/PhysRevLett.92.087902}{Experimental
  Detection of Multipartite Entanglement using Witness Operators},''
  \emph{Physical Review Letters}, vol.~92, p. 087902, 2004.

\bibitem{JMG11}
B.~Jungnitsch, T.~Moroder, and O.~G{\"u}hne,
  ``\href{https://link.aps.org/doi/10.1103/PhysRevA.84.032310}{Entanglement
  witnesses for graph states: General theory and examples},'' \emph{Physical
  Review A}, vol.~84, p. 032310, 2011.

\bibitem{ZZYM19}
Y.~Zhou, Q.~Zhao, X.~Yuan \emph{et~al.},
  ``\href{https://doi.org/10.1038/s41534-019-0200-9}{Detecting multipartite
  entanglement structure with minimal resources},'' \emph{npj Quantum
  Information}, vol.~5, no.~1, p.~83, 2019.

\bibitem{Bravyi21}
S.~Bravyi, S.~Sheldon, A.~Kandala \emph{et~al.},
  ``\href{https://link.aps.org/doi/10.1103/PhysRevA.103.042605}{Mitigating
  measurement errors in multiqubit experiments},'' \emph{Physical Review A},
  vol. 103, p. 042605, 2021.

\bibitem{M3}
P.~D. Nation, H.~Kang, N.~Sundaresan \emph{et~al.},
  ``\href{https://link.aps.org/doi/10.1103/PRXQuantum.2.040326}{Scalable
  mitigation of measurement errors on quantum computers},'' \emph{PRX Quantum},
  vol.~2, p. 040326, 2021.

\bibitem{Colin2022}
C.~K.-U. Becker, N.~Tcholtchev, I.-D. Gheorghe-Pop \emph{et~al.},
  ``\href{https://ieeexplore.ieee.org/document/9779849}{Towards a Quantum
  Benchmark Suite with Standardized {KPIs}},'' in \emph{2022 IEEE 19th
  International Conference on Software Architecture Companion (ICSA-C)}, 2022,
  pp. 160--163.

\bibitem{Ohkura2022}
Y.~Ohkura, T.~Satoh, and R.~Van~Meter,
  ``\href{https://ieeexplore.ieee.org/document/9749894}{Simultaneous Execution
  of Quantum Circuits on Current and Near-Future {NISQ} Systems},'' \emph{IEEE
  Transactions on Quantum Engineering}, vol.~3, pp. 1--10, 2022.

\bibitem{Niu2023}
S.~Niu and A.~Todri-Sanial,
  ``\href{https://doi.org/10.22331/q-2023-02-16-925}{Enabling Multi-programming
  Mechanism for Quantum Computing in the {NISQ} Era},'' \emph{{Quantum}},
  vol.~7, p. 925, 2023.

\bibitem{Liu2022}
L.~Liu and X.~Dou, ``\href{https://arxiv.org/abs/2207.14483}{{QuCloud+}: A
  Holistic Qubit Mapping Scheme for Single/Multi-programming on {2D/3D NISQ}
  Quantum Computers},'' \emph{arXiv preprint arXiv:2207.14483}, 2022.

\end{thebibliography}

\vfill


\end{document}